\documentclass{jfm}
\usepackage{graphicx}
\usepackage{amsbsy}
\usepackage{subfigure}
\usepackage[english]{babel}
\usepackage{color,amssymb,amsmath,verbatim,wasysym}
\usepackage{natbib, mathrsfs}
\usepackage{soul}

\usepackage{scalerel}
\usepackage{stackengine}
\setstackEOL{\#}
\stackMath
\def\hatgap{2pt}
\def\subdown{-2pt}
\newcommand\reallywidehat[2][]{%
\renewcommand\stackalignment{l}%
\stackon[\hatgap]{#2}{%
\stretchto{%
    \scalerel*[\widthof{$#2$}]{\kern-.6pt\bigwedge\kern-.6pt}%
    {\rule[-\textheight/2]{1ex}{\textheight}}
}{0.5ex}
_{\smash{\belowbaseline[\subdown]{\scriptstyle#1}}}%
}}

\newcommand{\mode}{h}

\newcommand*\mystrut[1]{\vrule width0pt height0pt depth#1\relax}

\newcommand{\rE}{\Upsilon}
\newcommand{\E}{\upsilon}
\newcommand{\kw}{k_0}
\newcommand{\dsigma}{\varsigma}
\newcommand{\bU}{\boldsymbol{U}}
\newcommand{\speed}{\mathcal{V}}
\newcommand{\pe}{e^B}
\newcommand{\ew}{e^A}
\newcommand{\pE}{\mathcal{E}^B}
\newcommand{\Ew}{\mathcal{E}^A}
\newcommand{\action}{\mathcal{A}}

\newcommand{\A}{A}

\newcommand{\com}{\, ,}
\newcommand{\per}{\, .}

\newcommand{\defn}{\ensuremath{\stackrel{\mathrm{def}}{=}}}

\def\beq{\begin{equation}}
\def\eeq{\end{equation}}

\newcommand{\disp}{\mathrm{D}}
\newcommand{\wave}{\mathrm{E}}

\newcommand{\pnabla}{\boldsymbol \nabla_{\! \! \perp}}
\newcommand{\hnabla}{\bnabla_{\! \! h}}
\newcommand{\bnablad}{\bnabla_{\! \! \alpha}}
\newcommand{\hlap}{\triangle}

\newcommand{\J}{\mathrm{J}}
\renewcommand{\L}{\mathrm{L}}
\newcommand{\M}{\mathrm{M}}

\newcommand{\D}{\mathrm{D}}

\newcommand{\bu}{\boldsymbol u}

\newcommand{\bx}{\boldsymbol x}

\newcommand{\bxh}{\hspace{0.1em} \boldsymbol{\hat x}}
\newcommand{\byh}{\hspace{0.1em}\boldsymbol{\hat y}}
\newcommand{\bzh}{\hspace{0.1em}\boldsymbol{\hat z}}

\newcommand{\ep}{\epsilon}

\newcommand{\ee}{\mathrm{e}}
\newcommand{\ii}{\mathrm{i}}
\newcommand{\cc}{\mathrm{cc}}

\newcommand{\dd}{{\rm d}}
\newcommand{\id}{{\, \rm d}}

\newcommand{\Bu}{Bu}

\newcommand{\half}{\tfrac{1}{2}}

\renewcommand{\tt}{\tilde t}
\newcommand{\bt}{\bar t}

\begin{document}

\title[Internal tides in quasi-geostrophic flow]{An asymptotic model for the propagation of oceanic internal tides through quasi-geostrophic flow}

\author[Wagner, Ferrando, and Young]{G.L. Wagner$^1$, G. Ferrando$^2$, and W.R. Young$^3$}

\affiliation{$^1$Department of Earth, Atmospheric, and Planetary Sciences, \\
Massachusetts Institute of Technology, Cambridge, Masschusetts 02139-4307, USA \\
$^2$D\'epartement de Physique, Ecole Normale Sup\'erieure, 24, rue Lhomond, 75005 Paris, France \\
$^3$Scripps Institution of Oceanography, University of California San Diego, \\ La Jolla, California 92093-0213, USA
}

\pubyear{}
\volume{}
\pagerange{}
\date{ \today}
\setcounter{page}{1}

\maketitle

\begin{abstract}

Starting from the hydrostatic Boussinesq equations, we derive a time-averaged `hydrostatic wave equation'  that describes the propagation of inertia-gravity internal waves through quasi-geostrophic flow.  The derivation uses a multiple-time-scale asymptotic method to isolate wave field evolution over intervals much longer than a wave period, assumes that the wave field has a well-defined and non-inertial frequency such as that of the mid-latitude semi-diurnal lunar tide, neglects nonlinear wave-wave interactions and makes no restriction on either the background density stratification or the relative spatial scales between the wave field and quasi-geostrophic flow.  As a result the hydrostatic wave equation is a reduced model applicable to the propagation of large scale internal tides through the inhomogeneous and moving ocean.  A numerical comparison with the linearized and hydrostatic Boussinesq equations demonstrates the validity of the hydrostatic wave equation and illustrates the manners of model failure when the quasi-geostrophic flow is too strong and the wave frequency is too close to inertial.

\end{abstract}

\section{Introduction}
\label{introduction}

Oceanic internal tides are inertia-gravity waves with tidal frequencies generated when tides slosh the rotating and stratified ocean over underwater hills and mountains.  While tides are planetary-scale surface waves forced by the gravitational pull of the sun and moon \citep{balmforth20052004}, internal tides are freely-propagating, subsurface internal waves with much smaller 100 km horizontal scales.  And while tides are forecast to within a centimeter in the open ocean, internal tides cannot be predicted at present due to their small scales and strong modulation by ever-changing eddies and currents \citep{rainville2006propagation,zaron2014time}.  

Internal tides are an energetic component of motion almost everywhere in the Earth's ocean.  Their strength and unpredictability means internal tides often provoke irritation by contaminating temporally-sparse data intended to observe more persistent flows \citep{wunsch1975internal, munk1981internal, ponte2015incoherent}. Such power betrays their intrinsic importance, too:  the terawatt or so that internal tides draw from surface tides \citep{egbert2000significant}  slows the spinning of the Earth, contributes to the outward drift of the moon and may drive the mixing and lifting of dense water that determines the ocean's density stratification. 

The explicit connection between internal tides and the evolution of oceanic stratification is unclear because the spatial distribution of internal tide energy and dissipation in small-scale ocean-mixing turbulence is not known.  Of the total energy transferred to internal tides, only a subordinate fraction of perhaps 8--40\% dissipates locally at generation sites, according to a small number of observational \citep{klymak2006estimate,alford2011energy,laurent2004examination} and model-based \citep{carter2008energetics} estimates.  The rest escapes into low-mode waves that propagate across ocean basins toward fates unknown.  Because the evolution of ocean stratification and circulation are sensitive to the horizontal and vertical distribution of tide-driven turbulent mixing \citep{melet2016climatic}, the long-range propagation and eventual dissipation of internal tides should be understood to ensure accurate prediction of the evolution of Earth's climate.

The primary obstacle to mapping and predicting the internal tide is the effect of inhomogeneous and time-varying oceanic flows on internal tide propagation \citep{rainville2006propagation,zaron2014time, ponte2015incoherent}.   For example, \cite{zaron2014time} conclude that horizontal density gradients associated with quasi-geostrophic flows are primarily responsible for scattering internal tides as they propagate away from the Hawaiian ridge.  This scattering process may extract energy from quasi-geostrophic flow, according to the thought experiment by \citet{buhler2005wave} and inferences drawn from the 1978-1979 POLYMODE Local Dynamics experiment by \cite{polzin2010mesoscale}.  Such a wave-flow interaction has implications both for large-scale dynamics as well as the energy available for wave-driven mixing.  The transfer of quasi-geostrophic energy into waves with near-inertial frequency manifests in the asymptotic theories of \cite{XieVanneste} and \cite{wagner2016near} and the simulations of \cite{barkan}. But the potential for transfer of quasi-geostrophic energy into waves with non-inertial frequencies has not been thoroughly explored.

The need to better understand interactions between internal tides and quasi-geostrophic flow motivates our asymptotic derivation of the `hydrostatic wave equation' exhibited in \eqref{internalTideEqnIntro}.  The hydrostatic wave equation describes the propagation of  three-dimensional, hydrostatic internal waves through a prescribed quasi-geostrophic mean flow and arbitrary density stratification.  The derivation uses a multiple time-scale asymptotic method to simplify and isolate the slow evolution of the wave field over time-scales much longer than a wave period.  The hydrostatic wave equation does not restrict the relative spatial scales of waves and flow and is thus applicable to oceanic scenarios in which internal tides and quasi-geostrophic flows coevolve on horizontal scales of 50 to 200 km \citep{chelton2011global, rocha2016mesoscale}.

The approximations used to derive the hydrostatic wave equation are intermediate between the reductions of geometric optics that permit ray tracing and the more mild assumptions of models linearized around arbitrary mean flows.  The ray tracing employed by \citet{rainville2006propagation} and conservation equations derived by \cite{salmon2016}, for example, require mean flows that vary on spatial scales much larger than a single wavelength.  This approximation is usually inappropriate for the low-mode oceanic internal tide.  On the other end of the spectrum of approximations is the `Coupled-mode Shallow Water' model developed by \cite{kelly}, which linearizes hydrostatic Boussinesq dynamics around a mean flow of arbitrary scale and strength.  The coupled-mode model is more general than the hydrostatic wave equation, but consists of three equations and resolves rapid oscillations on tidal frequencies.  In contrast, the hydrostatic wave equation is a single equation that filters tidal-frequency oscillations, but accepts only quasi-geostrophic mean flows.

The model that most resembles the hydrostatic wave equation is the spectral-space asymptotic model described by \citet{bartello1995geostrophic} and \citet{ward2010scattering}.  The derivation of this spectral model starts with the hydrostatic Boussinesq equations and assumes weak nonlinearity, so that the leading-order system describes linear wave propagation while the first-order system incorporates wave advection and refraction by quasi-geostrophic flow.  The wave field is then projected onto eigenfunctions or `wave modes' of the linear system and the resonant parts of the first-order system provide a set of ordinary differential equations that govern the evolution of each modal amplitude.  No assumption is made about the relative scales of waves and flow, but only exactly resonant interactions contribute to the evolution of a wave field that exactly satisfies the linear dispersion relation.  The hydrostatic wave equation relaxes this resonant interaction assumption, includes parts of the wave spectrum that do not exactly satisfy the linear dispersion relation and avoids the spectral decomposition through a physical-space `reconstitution' \citep{roberts1985introduction} of the leading- and first-order equations.  


\subsection{Summary of the hydrostatic wave equation}
\label{summary}

In the hydrostatic wave equation, the ocean's dynamic pressure field $p$ is decomposed into quasi-geostrophic and wave components so that
\beq
p = f_0 \left ( \psi + \ee^{- \ii \sigma t} A + \ee^{\ii \sigma t} A^* \right ) \com
\label{pressureIntro}
\eeq
where $\psi(x,y,z,t)$ is the quasi-geostrophic streamfunction, $A(x,y,z,t)$ is the complex amplitude of the wavy pressure field oscillating with frequency $\sigma$, and $f_0 = 4 \pi \sin \phi / \mathrm{day}$ is the constant local inertial frequency at latitude $\phi$.  Both $\psi$ and $A$ evolve slowly over time-scales much longer than $\sigma^{-1}$.  The pressure field in \eqref{pressureIntro} is a special solution to the rotating, hydrostatic Boussinesq equations that is justified only when initial conditions or oscillatory forcing select a combination of quasi-geostrophic and $\sigma$-frequency motion.  For the semidiurnal lunar tide $\sigma \approx 2 \pi \left ( 12.42 \, \, \mathrm{hours} \right )^{-1} \approx 1.405 \times 10^{-4} \, \mathrm{s^{-1}}$. 

The hydrostatic wave equation is derived by assuming that nonlinear interactions between $\psi$ and $A$ are small perturbations to linear balances in the hydrostatic Boussinesq equations \eqref{xmom} through \eqref{cont}.  The buoyancy field $b$ and velocity field $\bu = (u,v,w)$ are thus related to the hydrostatic pressure in \eqref{pressureIntro} through the linear hydrostatic Boussinesq equations and are given in equations \eqref{leadingOrderBuoyancy}, \eqref{leadingOrderVerticalVelocity} and \eqref{leadingOrderVelocity} below.

\citet{WagnerYoung} show that internal waves and quasi-geostrophic flow are distinguished by their imprint on available potential vorticity, or `APV'.  For small Rossby number flows the leading contribution to APV is $Q =N^2 \left [  v_y - u_x + \p_z \left ( f_0 b / N^2 \right ) \right ]$; and for linear waves $Q=0$, while for quasi-geostrophic flow APV is proportional to the classical quasi-geostrophic potential vorticity.  The requirement that $A$ in \eqref{pressureIntro} has negligible APV implies the approximate, dispersion-relation-like constraint 
\beq
\disp A \approx 0\com \label{dispersionConstraintIntro}
\eeq
where the `dispersion operator' $\disp$ is
\beq
\disp \defn  \underbrace{\mystrut{2.2ex} \p_x^2 + \p_y^2}_{\defn \hlap} \, - \, \alpha \, \underbrace{\mystrut{2.2ex} \p_z \frac{f_0^2}{N^2} \p_z}_{\defn \L} \per
\label{dispersionConstraintIntro1}
\eeq
In \eqref{dispersionConstraintIntro1} $N(z)$ is the buoyancy frequency at depth $z$ associated with an arbitrary background density stratification and we have defined the horizontal Laplacian $\hlap$ and vertical-derivative operator $\L$.  The non-dimensional parameter 
\beq
\alpha \defn \frac{\sigma^2 - f_0^2}{f_0^2}
\label{alphaDef}
\eeq
is the wave Burger number.  For a plane wave in density stratification with constant $N$, the constraint $\D A \approx 0$ implies that the $\sigma$-frequency wave approximately satisfies the linear hydrostatic dispersion relation and that $\alpha = \left ( N k / f_0 m \right )^2$ is its squared aspect ratio when $k \sim L^{-1}$ and $m \sim H^{-1}$ are horizontal and vertical wavenumbers.  The hydrostatic wave equation becomes invalid for quasi-geostrophic flow of a particular strength as $\alpha \to 0$ and the wave field becomes near-inertial.  Near-inertial waves are thus better described by the linearized YBJ equation of \citet{YBJ}, or the nonlinear models developed by \citet{XieVanneste} and \citet{wagner2016near}.  

The approximate equality in \eqref{dispersionConstraintIntro} would be exact if the wave field were constrained to have identically zero linear APV and thus exactly satisfy the linear dispersion relation.  The essence of our derivation is relax this constraint by `reconstituting' the leading-order equation, $\disp A = 0$, with the first-order equation that describes the nonlinear interaction of $\psi$ and $A$.  The result is a slow evolution equation for $A$, 
\beq
\begin{split}
 &\wave A_{t} + \J \left ( \psi , \wave A \right ) + \ii \alpha \sigma \disp A   + \J \left ( A, \disp \psi \right )  \\
& \qquad - \tfrac{2 \ii \sigma}{f_0^2} \Big [  \J \left ( \psi_x , \ii \sigma A_x - f_0 A_y \right ) +  \J \left ( \psi_y , \ii \sigma A_y + f_0 A_x \right ) \Big ] \\
& \qquad \qquad + \tfrac{ \ii \sigma}{f_0}  \left [ \hnabla \bcdot \left ( \disp \psi \hnabla A \right ) - \D \left ( \tfrac{\alpha f_0^2}{N^2} \psi_z A_z \right ) + \p_z \left ( \tfrac{\alpha f_0^2}{N^2} \psi_z \D A \right ) \right ]  = 0 \com
\end{split}
\label{internalTideEqnIntro}
\eeq
where $\hnabla \defn \p_x \bxh + \p_y \byh$ is the horizontal gradient,  the Jacobian  is $\J(a,b) = a_x b_y - a_y b_x$ and the operator $\wave$ is 
\beq
\wave \defn  \frac{\alpha}{2} \big [ \hlap + \left ( 4 + 3 \alpha \right ) \L \big ] \per
\eeq
The hydrostatic wave equation \eqref{internalTideEqnIntro} describes the slow evolution of hydrostatic internal waves with a pressure field given by \eqref{pressureIntro}, in three-dimensional quasi-geostrophic flow with streamfunction $\psi(x,y,z,t)$ of arbitrary spatial scale and non-uniform background stratification with buoyancy frequency $N(z)$.  

We begin the derivation of \eqref{internalTideEqnIntro} by non-dimensionalizing the hydrostatic Boussinesq equations and their associated `wave operator form' in section \ref{equationsTide}.  In section \ref{derivation} we use multiple-scale asymptotics and the method of reconstitution to derive a preliminary form of the hydrostatic wave equation.  In section \ref{remodeling} we make heuristic modifications that improve the result from section \ref{derivation} to finally arrive at \eqref{internalTideEqnIntro}.  In section \ref{nonconservation} we discuss a `non-conservation law' of \eqref{internalTideEqnIntro} that pertains to the classical conservation of energy and wave action.  We next define the region of validity of the hydrostatic wave equation by comparing 60 solutions to \eqref{internalTideEqnIntro} with the hydrostatic Boussinesq equations linearized around decaying two-dimensional turbulence in section \ref{validation}.  The comparison reveals how the model fails when the wave frequency is near-inertial or when the mean flow is too strong by considering a range of wave frequencies and turbulent mean flows.  We conclude by discussing the physical implications and potential applications of the hydrostatic wave equation and its relatives in section \ref{discussion}.

\section{The hydrostatic Boussinesq equations and  `wave operator form'}
\label{equationsTide}

The hydrostatic, rotating Boussinesq equations with constant inertial frequency $f = f_0$ are
\begin{align}
u_t + \bu \bcdot \bnabla u - f_0 v + p_x &= 0 \com \label{xmom} \\
v_t + \bu \bcdot \bnabla v + f_0 u + p_y &= 0 \com \label{ymom} \\
p_z &= b \com \label{zmom} \\
b_t + \bu \bcdot \bnabla b + w N^2 &= 0 \com \label{buoy} \\
u_x + v_y + w_z &= 0 \label{cont} \per
\end{align}
The hydrostatic approximation made in \eqref{zmom} is sensible for motions with large horizontal scales and small vertical scales, which implies that vertical velocities and vertical accelerations are relatively small.  For motions of frequency $\sigma$, the continuity equation \eqref{cont} and linear terms in the buoyancy equation \eqref{buoy} imply that
\beq
w \sim \frac{H}{L} u \qquad \text{and} \qquad b \sim \frac{N_0^2}{\sigma} w \sim \frac{N_0^2 H}{\sigma L} u \com
\label{buoyancyScaling}
\eeq
where $H$ and $L$ are the characteristic vertical and horizontal scales of the $\sigma$-frequency motion and $N_0$ is the characteristic magnitude of the buoyancy frequency profile $N(z)$.  In consequence, the assumption $w_t / b \ll 1$ underlying the hydrostatic approximation is valid for motions with frequency $\sigma$ when
\beq
\frac{w_t}{b} \sim \left ( \frac{\sigma}{N_0} \right )^{\! 2} \ll 1 \per
\label{hydrostaticCondition}
\eeq

Appendix \ref{waveOperatorFormAppendix} condenses the hydrostatic Boussinesq equations \eqref{xmom}--\eqref{cont} to their `wave operator form',
\beq
\begin{split}
\p_t \Big [ \p_t^2 \L + f_0^2 \left ( \hlap + \L \right ) \Big ] p &= - f_0^2 \left ( \p_t \hnabla + f_0 \pnabla \right )\bcdot \left ( \bu \bcdot \bnabla \right ) \bu \\
& \qquad \qquad - \p_z \tfrac{f_0^2}{N^2} \left ( \p_t^2 + f_0^2 \right ) \left ( \bu \bcdot \bnabla p_z \right ) \per
\end{split}
\label{waveOperatorFormTide}
\eeq
The operators $\hlap$ and $\L$ in \eqref{waveOperatorFormTide} are defined in \eqref{dispersionConstraintIntro1}, while
\beq
\hnabla = \p_x \bxh + \p_y \byh \qquad \text{and} \qquad     \pnabla   \defn  - \p_y \bxh + \p_x \byh \per
\eeq
The left side of \eqref{waveOperatorFormTide} is the hydrostatic internal wave operator acting on $p$, and the right-side collects the nonlinear terms. 

\subsection{Tidally-appropriate non-dimensionalization}

We non-dimensionalize the hydrostatic Boussinesq equations in \eqref{xmom}--\eqref{cont} by scaling $x,y$ with $L$ and $u,v$ with $U$ and thus assuming that both waves and quasi-geostrophic flow share length scales and velocity scales.  The emergent non-dimensional parameter
\beq
\ep \defn \frac{U}{f_0 L} 
\eeq
is then both the Rossby number and a measure of wave amplitude.  We assume $\ep \ll 1$ so that linear balances dominate \eqref{xmom}--\eqref{cont}.  

The $\sigma$-frequency `wave Burger number',
\beq
\alpha = \frac{\sigma^2 - f_0^2}{f_0^2} \com
\eeq
emerges as a critically important parameter.  Our use of common horizontal and vertical scales $L$ and $H$ implies that the Burger number of the quasi-geostrophic flow is 
\beq
\Bu \defn \left ( \frac{N_0 H}{f_0 L} \right )^{\! 2} \sim \alpha \per 
\eeq
The assumption $\Bu \sim \alpha = O(1)$ means we consider both wavy and quasi-geostrophic motions with aspect ratio $H/L \sim f_0 / N_0$.  The additional ocean-appropriate assumption $f_0/N_0 \ll 1$ implies that waves and flow have small aspect ratios with $H/ L \ll 1$ and permits the hydrostatic approximation in \eqref{zmom}. 

The wavy buoyancy scaling in \eqref{buoyancyScaling} and momentum equations \eqref{zmom} and \eqref{xmom} imply that
\beq
p \sim \frac{ \left ( N_0 H \right )^2 U}{\sigma L} \qquad \text{and} \qquad \frac{p_x}{u_t} \sim \left ( \frac{N_0 H}{\sigma L} \right )^{\! 2} = O(1) \per
\label{pressureScaling}
\eeq
We consider waves with $\alpha = O(1)$ and $\sigma / f_0 = O(1)$, so that $H / L \sim \sigma / N_0 \sim f_0 / N_0$.

Using these transformations, the hydrostatic Boussinesq equations \eqref{xmom}--\eqref{cont} become
\begin{align}
u_t - v + p_x &= - \ep \, \bu \bcdot \bnabla u \com \label{xmomTide} \\
v_t + u + p_y &= - \ep \, \bu \bcdot \bnabla v \com \label{ymomTide} \\
p_z - b &= 0 \com \label{zmomTide} \\
b_t + w N^2 &= - \ep \, \bu \bcdot \bnabla b \com \label{buoyTide} \\
u_x + v_y + w_z &= 0 \com \label{contTide}
\end{align}
while the wave operator form in \eqref{waveOperatorFormTide} becomes
\beq
\p_t \Big [ \p_t^2 \L +  \hlap + \L  \Big ] p = - \ep \, \Big [ \left ( \p_t \hnabla + \pnabla \right ) \bcdot \left ( \bu \bcdot \bnabla \right ) \bu +  \p_z \tfrac{1}{N^2} \left ( \p_t^2 + 1 \right ) \left ( \bu \bcdot \bnabla p_z \right ) \Big ] \per
\label{waveOperatorFormTideND}
\eeq
The derivation that follows in section \ref{derivation} expands \eqref{waveOperatorFormTideND} assuming that the right side is much smaller than the left.

\subsection{The two-time expansion}

To isolate the slow evolution of almost-linear waves over time-scales much longer than the fast time-scales of oscillation and linear dispersion, we propose the two-time expansion
\beq
\p_t \mapsto \p_{\tt} + \ep \, \p_{\bt} \com 
\label{twoTimeExpansion}
\eeq
where $\tt \sim f_0^{-1}$ is the fast time scale of wave oscillations and $\bt \sim L / U = \left ( \ep f_0 \right )^{-1}$ is the time-scale of slow wave evolution due to advection and refraction by quasi-geostrophic flow.  The two-time expansion in \eqref{twoTimeExpansion} transforms the wave operator in \eqref{waveOperatorFormTideND} into
\beq
\p_t \Big [ \p_t^2 \L + f_0^2 \left ( \hlap + \L \right ) \Big ] \mapsto \left ( \p_{\tt} + \ep \, \p_{\bt} \right ) \Big [ \left ( \p_{\tt} ^2 + 2 \ep \p_{\tt} \p_{\bt} + \ep^2 \p_{\bt} \right ) \L + f_0^2 \left ( \hlap + \L \right ) \Big ] \per
\label{twoTimeWaveOperator}
\eeq
The $O(1)$ terms in \eqref{twoTimeWaveOperator} comprise the linear Boussinesq wave operator
\beq
\p_{\tt} \Big [ \p_{\tt}^2 \L + f_0^2 \left ( \hlap + \L \right ) \Big ] \com
\eeq
while the $O(\ep)$ terms are
\beq
\ep \, \p_{\bt} \Big [ 3 \p_{\tt}^2 \L + f_0^2 \left ( \hlap + \L \right ) \Big ] \per
\label{o1WaveOperator}
\eeq
We do not write the two-timed form of the full Boussinesq system in \eqref{xmomTide}--\eqref{contTide} because we require only its linear, leading-order terms on the left side of each equation.  With the linear Boussinesq equations and the two-timed form of \eqref{waveOperatorFormTideND} we are ready to develop the asymptotic expansion that leads to the hydrostatic wave equation.  

\section{The hydrostatic wave equation}
\label{derivation}

We isolate the slow evolution of hydrostatic internal waves over the long time-scales of $\bar t$ by developing a perturbation expansion of both the hydrostatic Boussinesq equations \eqref{xmomTide}--\eqref{contTide} and their wave operator form \eqref{waveOperatorFormTideND} assuming that $\ep \ll 1$.  To this end we expand $\bu$, $b$, and $p$ in $\ep$, so that $p$ becomes, for example, 
\beq
p = p_0 + \ep \, p_1 + \cdots \per
\eeq
We develop \eqref{xmomTide}--\eqref{waveOperatorFormTideND} in orders of $\ep$ and express the result in dimensional variables for clarity.  

\subsection{At leading-order: linear dispersion and geostrophic balance}

The leading-order terms in the hydrostatic Boussinesq equations in \eqref{xmomTide}--\eqref{contTide} are 
\begin{align}
u_{0\tt} - f_0 v_0 + p_{0x} &= 0 \com \label{o0xmomTide} \\
v_{0\tt} + f_0 u_0 + p_{0y} &= 0 \com \label{o0ymomTide} \\
p_{0z} &= b_0 \com \label{o0zmomTide} \\
b_{0\tt} + w_0 N^2 &= 0 \com \label{o0buoyTide} \\
u_{0x} + v_{0y} + w_{0z} &= 0 \per \label{o0contTide}
\end{align}
while the leading-order terms from the wave operator equation \eqref{waveOperatorFormTideND} are
\beq
\p_{\tt} \Big [ \p_{\tt}^2 \L + f_0^2 \left ( \hlap + \L \right ) \Big ] p_0 = 0 \per
\label{leadingOrderWaveOperator}
\eeq
We assume the leading-order solution to \eqref{leadingOrderWaveOperator} can be written as the sum of a quasi-geostrophic streamfunction and a wave field with frequency $\sigma$, so that
\beq 
p_0 = f_0 \left (  \psi + \ee^{- \ii \sigma \tt} A_0 + \ee^{\ii \sigma \tt} A_0^* \right ) \per
 \label{leadingOrderPressure}
\eeq
Both $A_0$ and $\psi$ depend on $\bx$ and the slow time $\bt$ and have streamfunction units, so that $\pnabla A_0$ and $\pnabla \psi$ have units of velocity.  Equations \eqref{o0xmomTide} and \eqref{o0ymomTide} imply that $\psi$ obeys geostrophic balance.  

Equation \eqref{leadingOrderWaveOperator} implies that $A_0$ obeys the linear $\sigma$-frequency dispersion relation: 
\beq
- \ii \sigma f_0^3 \big [ \; \underbrace{\hlap - \alpha \L}_{= \disp} \; \big ] A_0 = 0 \com
\label{leadingOrderAEqn}
\eeq
where $\alpha = \sigma^2 / f_0^2 - 1$ is the wave Burger number and $\disp = \hlap - \alpha \L$ is the dispersion operator defined in \eqref{dispersionConstraintIntro1}.  When $\sigma = 2 f_0$, $\disp = \hlap - 3 \L$ is the operator that appears conspicuously in the $2f_0$ equation of \citet{wagner2016near}.

Equation \eqref{o0zmomTide} implies that
\beq
b_0 = f_0 \left ( \psi_z + \ee^{-\ii \sigma \tt} A_{0z} + \ee^{\ii \sigma \tt} A^*_{0z} \right ) \com
\label{leadingOrderBuoyancy}
\eeq
and \eqref{o0buoyTide} subsequently yields
\beq
w_0 = \frac{\ii \sigma f_0}{N^2} \left ( \ee^{-\ii \sigma \tt} A_{0z} - \ee^{\ii \sigma \tt} A^*_{0z} \right ) \per
\label{leadingOrderVerticalVelocity}
\eeq
By merging $\p_{\tt} \eqref{o0xmomTide} + f_0 \eqref{o0ymomTide}$ with $\p_{\tt} \eqref{o0ymomTide} - f_0\eqref{o0xmomTide}$ we obtain a single vector equation for horizontal velocity $\bu_{0h} = u_0 
\bxh + v_0 \byh $,
\beq
\left ( \p_{\tt}^2 + f_0^2 \right ) \bu_{0h} = - \big ( \p_{\tt} \hnabla - f_0 \pnabla \big ) p_0 \com
\eeq
which we solve given $p_0$ in \eqref{leadingOrderPressure}.  The three velocity components are then
\begin{align}
\left ( \begin{matrix}
u_0 \\
v_0 \\
w_0 
\end{matrix} \right ) 
&= 
\left ( \begin{matrix}
- \p_y \ \\
\p_x \\
0 
\end{matrix} \right ) \psi
- 
 \frac{1}{\alpha f_0} \left ( \begin{matrix}
 \ii \sigma \p_x - f_0 \p_y   \\[1ex]
 \ii \sigma \p_y + f_0 \p_x   \\[1ex]
- \frac{\ii \sigma \alpha f_0^2 }{N^2} \p_z
\end{matrix} \right ) \ee^{-\ii \sigma \tt}  A_0
+
 \frac{1}{\alpha f_0} \left ( \begin{matrix}
 \ii \sigma \p_x + f_0 \p_y   \\[1ex]
  \ii \sigma \p_y - f_0 \p_x  \\[1ex]
- \frac{\ii \sigma \alpha f_0^2}{N^2} \p_z
\end{matrix} \right ) \ee^{\ii \sigma \tt}  A_0^* \com 
\label{leadingOrderVelocity}
\end{align}
where we have used  $\sigma^2 - f_0^2 = \alpha f_0^2$.  More properties of the leading-order solution to \eqref{o0xmomTide}--\eqref{o0contTide} are given in appendix \ref{RHSleadingOrderSolution}.

\subsection{At first-order: slow wave evolution}

The $O(\ep)$ terms in the wave operator equation \eqref{waveOperatorFormTideND} are
\begin{align}
\begin{split}
f_0^3 \left ( \hlap + \L \right ) \psi_{\bt} \, + \, &  f_0 \Big [ f_0^2 \hlap - \left ( 3 \sigma^2 - f_0^2 \right ) \L \Big ]  \Big [ \ee^{- \ii \sigma \tt} \! A_{0\bt} + \ee^{\ii \sigma \tt} A^*_{0\bt} \Big ] \\
& \qquad \qquad \qquad + \Big [ \p_{\tt}^2 \L + f_0^2 \left ( \hlap + \L \right ) \Big ] p_{1\tt} = \mathrm{RHS}(\psi, A_0) \per 
\label{o1WaveOperatorForm}
\end{split} 
\end{align}
In \eqref{o1WaveOperatorForm}, $\mathrm{RHS}(\psi, A_0)$ is short for the $O(\ep)$ nonlinear terms in \eqref{waveOperatorFormTideND} evaluated using the leading--order solution and defined by
\beq
\mathrm{RHS} \left ( \psi, A_0 \right ) \defn - f_0^2 \left ( \p_{\tt} \hnabla + f_0 \pnabla \right ) \bcdot \left ( \bu_0 \bcdot \bnabla \right ) \bu_0  - \p_z \frac{f_0^2}{N^2} \left ( \p_{\tt}^2 + f_0^2 \right ) \left ( \bu_0 \bcdot \bnabla b_0 \right ) \per
\label{RHSdef}
\eeq
Equation \eqref{o1WaveOperatorForm} describes the slow evolution and propagation of $A_0$, quasi-geostrophic evolution, and nonlinear wave dynamics that generate both quasi-steady mean flows and wave harmonics with frequency $2 \sigma$.

The quasi-geostrophic streamfunction $\psi$ evolves due to its advection of quasi-geostrophic potential vorticity, $q$, 
\beq
q_{\bar t} + \J \left ( \psi, q \right ) = 0 \com \qquad \text{where} \qquad q \defn \left ( \hlap + \L \right ) \psi \per
\label{theQGEquation}
\eeq
The fact that the quasi-geostrophic potential vorticity evolves independently of the wave field $A$ in \eqref{theQGEquation} is a consequence of the assumption that waves and flow share the common velocity scale $U$ and length scales $H$ and $L$.  The derivation of \eqref{theQGEquation} in the presence of a wave field is given by  \cite{bartello1995geostrophic} using an eigenfunction decomposition and in the introduction of \cite{wagnerThesis} using available potential vorticity. 

We focus on the slow evolution of $\sigma$-frequency motions by multiplying \eqref{o1WaveOperatorForm} with $\ee^{\ii \sigma \tt}$ and averaging the result in $\tt$ over a wave period $2 \pi / \sigma$.  The average is denoted with an overbar and defined by
\beq
\bar \phi(\bt) = \frac{\sigma}{2 \pi} \int_{\bt - \frac{\pi}{\sigma}}^{\bt + \frac{\pi}{\sigma}} \! \phi(\bt, \tt) \, \id \tt \com
\eeq
for any quantity $\phi(\bt, \tt)$.  This average has the property that $\overline{\ee^{2 \ii \sigma \tt} A_0^*} = 0$ and $\bar{A_0} = A_0$, for example, because $A_0$ does not depend on the fast time $\tt$.  In consequence, the operation $ \overline{ \ee^{\ii \sigma \tt} \eqref{o1WaveOperatorForm}}$ isolates terms in \eqref{o1WaveOperatorForm} proportional to $\ee^{-\ii \sigma \tt}$, yielding
\beq
f_0 \Big [ f_0^2 \hlap - \left ( 3 \sigma^2 - f_0^2 \right ) \L \Big ] A_{0 \bt} - \ii \sigma f_0^3 \disp A_1 = \overline{ \ee^{\ii \sigma \tt} \text{RHS} } \per
\label{crudeAsymptotics}
\eeq
In forming \eqref{crudeAsymptotics} we assume that $p_1$ takes the form
\beq
p_1 = f_0 \left ( \ee^{-\ii \sigma \tt} A_1 + \ee^{\ii \sigma \tt} A_1^* \right ) + \cdots \com
\eeq
where the dots represent unimportant steady and $2 \sigma$-frequency parts of $p_1$, and $A_1 = f_0^{-1} \overline{ \ee^{\ii \sigma \tt} p_1}$ is the $O(\ep)$ correction to $A_0$.

The bookkeeping required to parse RHS for terms proportional to $\ee^{-\ii \sigma \tt}$ and thus identify the right side of \eqref{crudeAsymptotics} is detailed in appendix \ref{findingRHS}.  After multiplying by $\alpha / f_0$ for presentation, the result is 
\beq
\begin{split}
\tfrac{\alpha}{f_0} \overline{ \ee^{\ii \sigma \tt} \text{RHS} } &= \left ( \sigma^2 + f_0^2 \right ) \J \left ( \psi, \hlap A_0 \right ) + \left ( \alpha f_0 \right )^2 \J \left ( \psi, \L A_0 \right ) + f_0^2 \J \left ( A_0, \disp \psi \right )   \\
& \quad - 2 \ii \sigma \Big [  \J \left ( \psi_x , \ii \sigma A_{0x} - f_0 A_{0y} \right ) +  \J \left ( \psi_y , \ii \sigma A_{0y} + f_0 A_{0x} \right ) \Big ] \\
& \qquad +  \ii \sigma f_0 \left [ \hnabla \bcdot \left ( \disp \psi \hnabla A_0 \right ) - \disp \left ( \tfrac{\alpha f_0^2}{N^2} \psi_z A_{0z} \right ) + \p_z \left ( \tfrac{\alpha f_0^2}{N^2} \psi_z \disp A_0 \right )  \right ] \per
\end{split}
\label{foundRHS}
\eeq
With \eqref{foundRHS} the major algebraic challenge in deriving the hydrostatic wave equation is behind us.  

Two different approaches may now be used to develop a wave evolution model from the leading-order equation \eqref{leadingOrderAEqn} and first-order equation \eqref{crudeAsymptotics}.  One approach is to move into the spectral space associated with eigenfunctions or `wave modes' of the operator $\disp$.  In this approach the first step is then to project the leading-order equation \eqref{leadingOrderAEqn} onto wave modes, which defines the spectral components of $A_0$ and solves \eqref{leadingOrderAEqn} exactly.  Next, projecting the first-order equation \eqref{crudeAsymptotics} onto wave modes eliminates $\disp A_1$ and isolates the slow evolution of those spectral components of $A_0$.  This strategy was employed, for example, by \cite{ward2010scattering} and \cite{bartello1995geostrophic}.  We take the second approach, however: reconstitution.

\subsection{Reconstitution}

Rather than solve the leading-order equation \eqref{leadingOrderAEqn} exactly, we instead reconstitute the asymptotic expansion by adding \eqref{leadingOrderAEqn} to the first-order equation \eqref{crudeAsymptotics} to obtain a single equation for the total wave amplitude $A = A_0 + A_1$.  After multiplying by $\alpha / f_0$ and rearranging terms, the result is
\beq
\begin{split}
& - \alpha \Big [ f_0^2 \hlap - \left ( 3 \sigma^2 - f_0^2 \right ) \L \Big ] A_{\bt} + \ii \alpha \sigma f_0^2 \disp A \\
& \quad + \left ( \sigma^2 + f_0^2 \right ) \J \left ( \psi, \hlap A \right ) + \left ( \alpha f_0 \right )^2 \J \left ( \psi, \L A \right ) + f_0^2 \J \left ( A, \disp \psi \right ) \\
& \quad \; \;- 2 \ii \sigma \Big [  \J \left ( \psi_x , \ii \sigma A_{x} - f_0 A_{y} \right ) +  \J \left ( \psi_y , \ii \sigma A_{y} + f_0 A_{x} \right ) \Big ] \\
& \; \qquad + \ii \sigma f_0 \left [ \hnabla \bcdot \left ( \disp \psi \hnabla A \right ) - \disp \left ( \tfrac{\alpha f_0^2}{N^2} \psi_z A_{z} \right ) + \p_z \left ( \tfrac{\alpha f_0^2}{N^2} \psi_z \disp A \right )  \right ] = O(\ep^3 f_0^4)  \per
\end{split}
\label{cruderAsymptotics}
\eeq
Excepting those that involve $\D A$, all terms on the left side of \eqref{cruderAsymptotics} scale with $\alpha f_0^2 \hlap A_{\bt} \sim \ep^2 f_0^4$.  The residual on the right side of \eqref{cruderAsymptotics} thus implies the error incurred during reconstitution is $O(\ep)$ and of same magnitude as terms already neglected by the perturbation expansion.  In this sense, \eqref{cruderAsymptotics} is asymptotically equivalent to the original hydrostatic Boussinesq equations.

One consequence of reconstitution is that the leading-order equation \eqref{leadingOrderAEqn} is not exactly satisfied so that $\D A \ne 0$ in general.  As a result, \eqref{cruderAsymptotics} describes the evolution of wave modes with frequencies slightly different than $\sigma$; or in other words, \eqref{cruderAsymptotics} describes both resonant and near-resonant interactions between $\psi$ and $A$.  On the other hand, because the dispersion terms $\ii \alpha f_0^2 \hlap A$ and $\ii \alpha^2 f_0^2 \L A$ in $\ii \alpha f_0^2 \D A$ are the largest in \eqref{cruderAsymptotics} by $\ep^{-1}$, solutions to \eqref{cruderAsymptotics} still approximately satisfy $\disp A \approx 0$ so that $A$ remains tethered to the $\sigma$-frequency hydrostatic dispersion relation when $\ep \ll 1$ and $\alpha = O(1)$.

\section{Remodeling}
\label{remodeling}

In principle, equation \eqref{cruderAsymptotics} achieves the goal of this paper and provides a valid description of the propagation of hydrostatic waves through quasi-geostrophic flows.  Yet several shortcomings either limit the range of its validity or prevent its practical implementation.  Its principal shortcoming is that the operator acting on $A_{\bar t}$ on the first line of \eqref{cruderAsymptotics} cannot be inverted in general.  The second shortcoming is that \eqref{cruderAsymptotics} is not Galilean invariant: its form is not preserved under translation by a uniform velocity implied by the two transformations $\psi \mapsto - U y + V x + \psi$ and $\p_{\bt} + U \p_x + V \p_y \mapsto \p_{\bt}$.  The lack of Galilean invariance hampers \eqref{cruderAsymptotics}'s description of Doppler shifting of wave field frequency by relatively uniform quasi-geostrophic flow.

We address these issues by modifying the model by adding two small $O(\ep^3 f_0^4)$ terms proportional to $\disp A_{\bt}$ and $\J \left ( \psi, \disp A \right )$ to \eqref{cruderAsymptotics}.  Formally, these two terms have the same magnitude as the error incurred in constructing \eqref{cruderAsymptotics}  and thus do not change the residual on the right of \eqref{cruderAsymptotics}.  Yet the judicious choice of proportionality significantly improves \eqref{cruderAsymptotics}'s approximation of linear wave dispersion and restores Galilean invariance.  

\subsection{An improved approximation to linear dispersion \label{subsecfourone}}

We first modify \eqref{cruderAsymptotics} by adding the linear term $c \alpha f_0^2 \disp A_{\bt}$, where $c$ is a constant determined by fitting the dispersion relation of the resulting equation to the exact dispersion relation implied by the hydrostatic Boussinesq system.  This improvement to \eqref{cruderAsymptotics} produces an equation that more faithfully describes exact linear dispersion when the spectrum of the wave field deviates from the wavenumber combinations $k^2 + \ell^2 = \alpha \kappa_n^2$.  

After dividing by $\alpha$, the linear terms in the modified equation $ \eqref{cruderAsymptotics} + c \alpha f_0^2 \disp A_{\bt}$ that remain when $\psi=0$  are
\beq
\left [ f_0^2 \left ( c + 1 \right ) \D A - 2  \sigma^2 \L \right ] A_{\bt} + \ii \sigma f_0^2 \left ( \hlap - \alpha \L \right ) A = 0 \per
\label{linearCruderTerms}
\eeq
Assuming the spectral representation $A \sim \ee^{\ii k x - \ii \dsigma \bar t} \mode_n(z)$, where $k$ is a horizontal wavenumber, $\dsigma$ is the deviation in wave frequency from $\sigma$, and $\mode_n$ are the hydrostatic vertical modes that solve the eigenproblem
\beq
\L \mode_n + \kappa_n^2 \mode_n = 0 \com \qquad \text{with} \qquad \mode_{nz} = 0 \quad \text{at} \quad z = -H, 0 \com
\label{verticalModesTide}
\eeq
leads to the linear dispersion relation implied by \eqref{linearCruderTerms}, 
\beq
\sigma + \dsigma = \sigma + \frac{ \sigma f_0^2 \left ( k^2 - \alpha \kappa_n^2 \right ) }{2 \left ( \sigma \kappa_n \right )^2 - f_0^2(c+1)\left ( k^2 - \alpha \kappa_n^2 \right ) } \per
\label{crudeDispRelation}
\eeq
The dispersion relation in \eqref{crudeDispRelation} is an expansion of the exact vertical mode-$n$ hydrostatic dispersion relation, 
\beq
\Sigma = \pm f_0 \sqrt{ 1 + \frac{k^2}{\kappa_n^2} } \com
\label{exactDispRelation}
\eeq
around the wavenumber combinations $k = \kappa_n \sqrt{\alpha}$ that correspond to $\Sigma = \sigma$.  

\begin{figure}
\centering
\includegraphics[width = 1\textwidth]{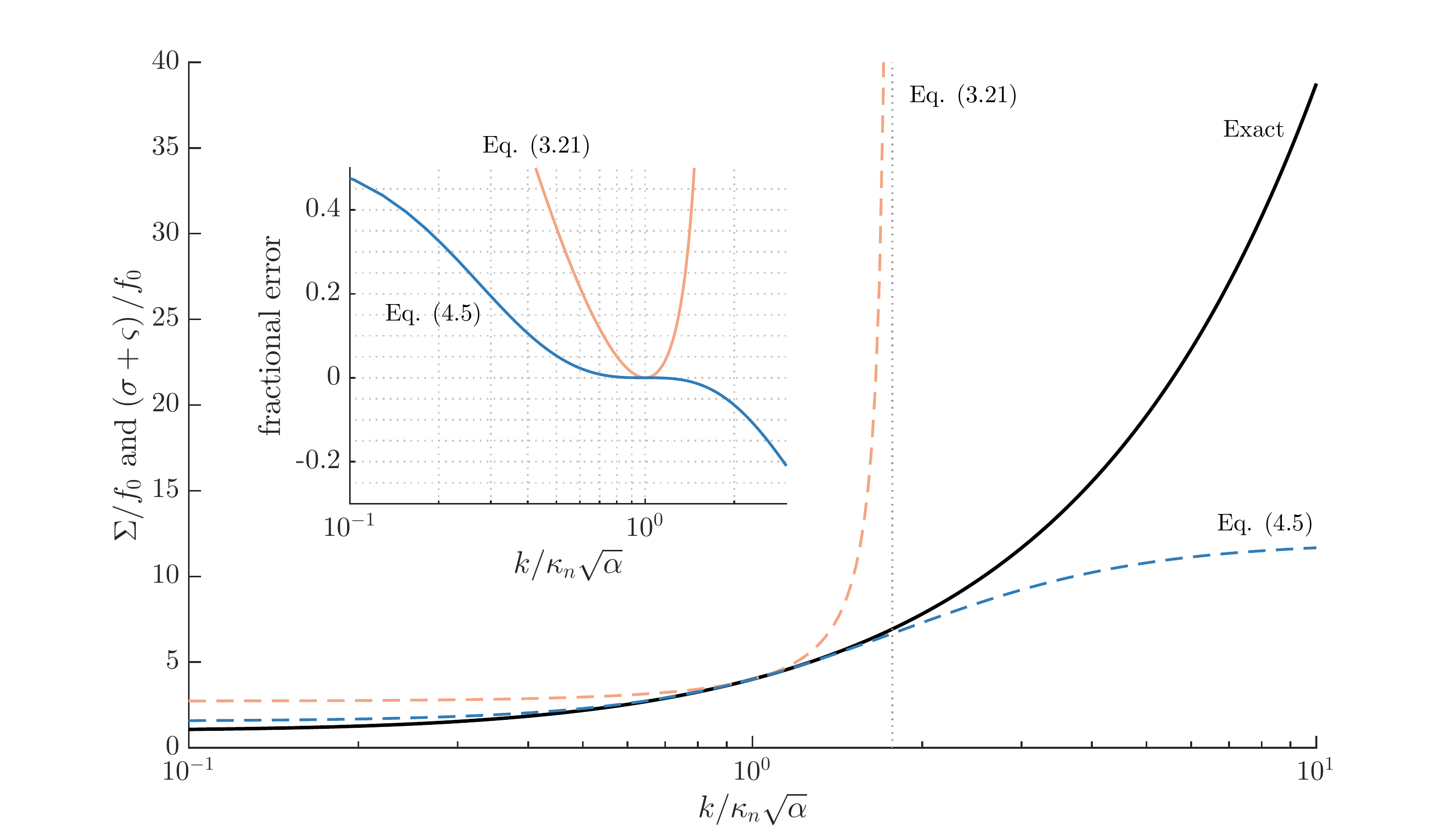}
\caption{Comparison of the exact hydrostatic mode-$n$ Boussinesq dispersion relation $\Sigma / f_0$ with the dispersion relations $\left ( \sigma + \dsigma \right ) / f_0$ implied by \eqref{cruderAsymptotics} and \eqref{crudererAsymptotics}.  $\Sigma / f_0$ is given in \eqref{exactDispRelation} while $\left ( \sigma + \dsigma \right ) / f_0$ for \eqref{cruderAsymptotics} and \eqref{crudererAsymptotics} are given by \eqref{crudeDispRelation} with $c=0$ and $c=-3/2$, respectively.  All three are plotted in the main figure versus $k / \kappa_n \sqrt{\alpha}$ on a logarithmic $x$-axis.  A gray dotted line shows the asymptote at $k / \kappa_n \sqrt{\alpha} = \sqrt{2\left ( 1 + \alpha^{-1} \right )}$ where the dispersion relation implied by \eqref{cruderAsymptotics} is undefined.  The inset shows the fractional error $\left ( \sigma + \dsigma - \Sigma \right ) / \Sigma$ versus $k / \kappa_n \sqrt{\alpha}$.}
\label{dispersionRelationComparison}
\end{figure}

Taking one derivative of \eqref{crudeDispRelation} and \eqref{exactDispRelation} with respect to $k$ while holding $\kappa_n$ constant reveals that $\Sigma_k = \dsigma_k$ at $k = \kappa_n \sqrt{\alpha}$. This means that  \eqref{linearCruderTerms} correctly captures the group velocity of waves with frequency $\sigma$ regardless of the value of $c$.  We choose $c = -3/2$, therefore, to match the second derivatives $\dsigma_{kk}$ and $\Sigma_{kk}$ so that the approximate dispersion relation $\sigma + \dsigma$ osculates the exact dispersion relation $\Sigma$.   The choice $c=-3/2$ also fixes the non-invertability of the operator that acts on $A_{\bar t}$ in \eqref{cruderAsymptotics}.  The linear terms in the improved equation $f_0^{-2} \eqref{cruderAsymptotics}  - 3 \alpha \D A_{\bt} / 2 $ that remain when $\psi=0$ are then 
\beq
\frac{\alpha}{2} \Big [ \hlap + \left ( 4 + 3 \alpha \right ) \L \Big ] A_{\bt} + \ii \alpha \sigma \D A = 0 \per
\label{crudererAsymptotics}
\eeq
Figure \ref{dispersionRelationComparison} compares the raw dispersion relation implied by \eqref{cruderAsymptotics} and the improved dispersion relation implied by \eqref{crudererAsymptotics} with the exact dispersion relation of the hydrostatic Boussinesq system. 

\subsection{Restoration of Galilean invariance \label{subsecfourtwo}}

The advection terms in \eqref{cruderAsymptotics} have the form $\J \left ( \psi , \bullet \right )$. These $\psi$-dependent terms remain when the mean flow $U \bxh + V \byh$ is horizontal and uniform with streamfunction $\psi = - U y + V x$.  Using $\sigma^2 / f_0^2 = \alpha + 1$, the advection terms in the improved equation $f_0^{-2} \eqref{cruderAsymptotics} - 3 \alpha \D A_{\bt} / 2$ become
\beq
\frac{\alpha}{2} \Big [ \hlap + \left ( 4 + 3 \alpha \right ) \L \Big ] A_{\bt} + \J \left ( \psi , \left [ \alpha + 2 \right ] \hlap A + \alpha^2 \L A \right ) \per
\label{advectionTerms}
\eeq
Remarkably, adding the small term
\beq
- \left ( \half + \tfrac{2}{\alpha} \right ) \J \left ( \psi, \disp A \right ) 
\eeq
to \eqref{advectionTerms} produces
\beq
\wave A_{\bt} + \J \left ( \psi, \wave A \right ) \com 
\label{advectionTerms2}
\eeq
where
\beq
\wave \defn \frac{\alpha}{2} \Big [ \hlap + \left ( 4 + 3 \alpha \right ) \L \Big ] \per
\label{advectionTerms3}
\eeq
The terms in \eqref{advectionTerms2} describe the advection of the wave quantity $\wave A$ by a velocity field associated with $\psi$.  Galilean invariance follows from the preservation of form under the simultaneous transformation $\psi \mapsto - U y + Vx + \psi$ and $\p_{\bt} \mapsto \p_{\bt} - U \p_x - V \p_y $.

The two remodeling steps produce the much improved equation 
\beq
f_0^{-2} \eqref{cruderAsymptotics} - \tfrac{3\alpha }{2} \D A_{\bt} - \left ( \half + \tfrac{2}{\alpha} \right ) \J \left ( \psi, \D A \right ) \com
\eeq
which rearranges into
\beq
\begin{split}
& \wave A_{\bt} + \J \left ( \psi , \wave A \right ) + \ii \alpha \sigma \disp A  + \J \left ( A, \disp \psi \right )  \\
& \qquad - \tfrac{2 \ii \sigma}{f_0^2} \Big [  \J \left ( \psi_x , \ii \sigma A_x - f_0 A_y \right ) +  \J \left ( \psi_y , \ii \sigma A_y + f_0 A_x \right ) \Big ] \\
& \qquad \qquad + \tfrac{ \ii \sigma}{f_0}  \left [ \hnabla \bcdot \left ( \disp \psi \hnabla A \right ) - \D \left ( \tfrac{\alpha f_0^2}{N^2} \psi_z A_z \right ) + \p_z \left ( \tfrac{\alpha f_0^2}{N^2} \psi_z \D A \right ) \right ]  =0 \per
\end{split}
\label{internalTideEqnRemodeling}
\eeq
For the final remodeling step, we drop the bar over $\bar t$ to write \eqref{internalTideEqnRemodeling} in terms of the single time-scale $t$.  The result is equation \eqref{internalTideEqnIntro}.

\subsection{Quasi-geostrophic perturbation of the mean stratification}

In sections \ref{subsecfourone} and \ref{subsecfourtwo} we added small terms to \eqref{cruderAsymptotics} produce the improved equation  \eqref{internalTideEqnRemodeling}.  Note, however, that \eqref{cruderAsymptotics} already contains one small term,
\beq
\p_z \left ( \tfrac{\alpha f_0^2}{N^2} \psi_z \D A \right ) \com \label{billzTerm}
\eeq
which has the same magnitude as terms neglected in constructing \eqref{cruderAsymptotics}.  We retain the small term \eqref{billzTerm} because it means the remodeled equation \eqref{internalTideEqnRemodeling} more faithfully encodes dynamics associated with a quasi-geostrophic perturbation to the background density stratification.  

This physical process is isolated by considering the case where $\psi(z)$ depends only on $z$, so that $\psi$ has no associated flow and acts only to perturb the buoyancy frequency from $N^2$ to $N^2 + f_0 \psi_{zz}$.  In this case the familiar vertical differential operator $\L$ defined in \eqref{dispersionConstraintIntro1} is correspondingly perturbed into
\beq
\p_z \frac{f_0^2}{N^2+ f_0\psi_{zz}} \p_z = \L \underbrace{- \,\p_z \frac{f_0^4}{N^4} \frac{\psi_{zz}}{f_0} \p_z }_{\defn \M} + \, O(\ep^2 \L )\com
\eeq
where $\M$ is the $O(\ep)$ perturbation to $\L$.  Similar to the principle that our model should retain the Boussinesq property of Galilean invariance in the case of uniform flow, a righteous approximation must capture the $O(\ep)$ perturbation to the density stratification and dispersion relation induced by $f_0 \psi_z$ and $\M$. 

Now consider the simplification of \eqref{internalTideEqnRemodeling} when $\psi = \psi(z)$.  First, the Jacobians on the first and second lines of \eqref{internalTideEqnRemodeling} all reduce to zero.  Next, some intricate simplifications of the third line of \eqref{internalTideEqnRemodeling}, aided by the non-obvious identity
\beq
\p_z \left ( \tfrac{f_0^2}{N^2} \psi_z \L A \right ) - \L \left ( \tfrac{f_0^2}{N^2} \psi_z A_z \right ) = f_0 \M A \com
\eeq
eventually reduce \eqref{internalTideEqnRemodeling} to 
\beq
\wave A_{\bt} + \ii \alpha \sigma \left[ \hlap  - \alpha \left(\L + \M \right)\right] A=0 \per 
\label{intricateSimp}
\eeq
The effect of the static streamfunction $\psi(z)$ is reduced to a transformation of the dispersion operator $\disp$ from $\hlap - \alpha \L$ to $\hlap - \alpha \left ( \L + \M \right )$.  The formation of the proper perturbed operator $\L+\M$ in \eqref{intricateSimp} requires the participation of the small term \eqref{billzTerm}.  The inclusion of \eqref{billzTerm} thus gives \eqref{internalTideEqnRemodeling} a more faithful description of the modification of internal wave dispersion by quasi-geostrophic perturbations to the density stratification.

\section{The non-conservation of wave action}
\label{nonconservation}

\cite{bretherton1968wavetrains} show that the amplitude of slowly-varying waves in inhomogeneous moving media is determined by the conservation of an adiabatic invariant called `wave action'.  Wave action is defined as wave energy divided by intrinsic frequency, or the frequency of the wave field measured by an observer moving with the local velocity of the medium.  Wave action conservation shows explicitly that wave field spatial distortions and associated shifts in frequency and spectral content are attended by transfers of energy with the inhomogeneous medium through which the waves propagate.

We ask whether a form of wave action is conserved by the hydrostatic wave equation \eqref{internalTideEqnIntro}, in which case the medium is a quasi-geostrophic flow that evolves slowly in time but varies rapidly in space.  For example, when the quasi-geostrophic flow varies slowly in both time and space, wave action is conserved \citep{salmon2016} and is used by \cite{buhler2005wave} to demonstrate that wave capture transfers quasi-geostrophic energy to the ocean's internal wave field.  
Also, the near-inertial equation derived by \cite{YBJ}, which is similar to equation \eqref{internalTideEqnIntro} above, conserves a form of wave action equal to the volume-integrated wave field kinetic energy divided by the local inertial frequency.   

In this section we show that equation \eqref{internalTideEqnIntro} does not conserve wave action.  Instead, \eqref{internalTideEqnIntro}'s version of wave action, which is similar but not equivalent to wave energy divided by its near-constant frequency $\sigma$, evolves as a direct consequence of wave field's non-satisfaction of the linear equations and associated non-adherence to a linear dispersion relation.  The inhomogeneity that forces wave action evolution originates from the term describing wave field advection by the non-uniform quasi-geostrophic flow.

An evolution equation for wave action in the hydrostatic wave equation emerges from the combination
\beq
\frac{1}{\alpha^2 \sigma} \int \! A^* \times \eqref{internalTideEqnIntro} + A \times\eqref{internalTideEqnIntro}^* \id V \com
\label{almostWaveAction}
\eeq
assuming that exact derivatives over the domain $V$ integrate to zero.  One useful identity that helps to simplify \eqref{almostWaveAction} writes the operator $\wave$ in terms of $\disp$, 
\beq
\wave = 2(1+\alpha) \hlap - \tfrac{4+3 \alpha}{2} \, \disp\com
\eeq
and a second forms an exact derivative from one of the horizontal refraction terms in \eqref{almostWaveAction},
\begin{equation}
A^*\hnabla\bcdot\left(\D\psi \, \hnabla A\right) - \A\hnabla\bcdot\left(\D\psi\, \hnabla A^*\right) = \hnabla\bcdot\left[ \D\psi\, \left(A^*\hnabla A - A\hnabla A^*\right)\right] \per
\end{equation}
A third identity, that leads to a cancellation between two Jacobians and part of the advection term $\J \left ( \psi, \wave A \right )$, is
\begin{align}
\int \!  A^* \J(\psi,\hlap A) + A \J(\psi,\hlap A^*) \id V
  = -\int \!  A^*\big [\J(\psi_x, A_x) + \J(\psi_y, A_y)\big]  \id V + \cc \per
\end{align}
Finally, we note that all the terms in \eqref{internalTideEqnIntro} with $\ii$ as factor cancel each other during the integration in \eqref{almostWaveAction}.  For example, a few integrations by parts yields the identity
\begin{align}
\int A^*\J(\psi_x,  A_y) - A^*\J(\psi_y, A_x)\, \dd V &= -\int\p_y \left [ A \J(\psi_x,A^*) \right ] - \p_x \left [ A\J(\psi_y,A^*) \right ] \id V \com 
\label{both1}\\
&= \int A \, \J(\psi_x,A_y^*) - A \, \J(\psi_y, A_x^*) \id V \per \label{both3}
\end{align}
Because \eqref{both3} is the complex conjugate of the left side of \eqref{both1}, both quantities are real and cancel during the simplification of \eqref{almostWaveAction}.

Assembling these and additional identities and using many integrations by parts eventually produces an evolution equation for $\action$, the wave action: 
\beq
\frac{\dd \action}{\dd t} = \frac{ 4 + 3 \alpha }{2 \alpha^2 \sigma } \int \psi \big [ \, \J \left ( A^*, \disp A \right ) + \J \left ( A, \disp A^* \right ) \big ] \id V \com
\label{actionNonConservation}
\eeq
where
\beq
\action \defn \frac{1}{2 \alpha \sigma} \int | \hnabla A |^2 + \left ( 4 + 3 \alpha \right ) \frac{f_0^2}{N^2} | A_z |^2 \id V \per
\label{almostAction}
\eeq
The magnitude of the residual on the right of \eqref{actionNonConservation} depends explicitly on the fact that $\disp A \ne 0$.  The residual on the right of \eqref{actionNonConservation} is smaller than the individual contributions on the left of \eqref{actionNonConservation} by $O(\ep)$.

The wave action in \eqref{almostAction} resembles, but is not equal to, Bretherton \& Garrett's definition of wave energy divided by intrinsic frequency.  The wave energy, or the wave-associated part of horizontal kinetic plus potential energy contained in the leading-order solution \eqref{leadingOrderBuoyancy} and \eqref{leadingOrderVelocity}, is defined in \eqref{perturbationEnergy} and given by
\beq
\Ew = \int \frac{\alpha + 2}{\alpha^2} | \hnabla A |^2 + \frac{f_0^2}{N^2} | A_z |^2 \id V \per
\eeq
Subtracting $ \left ( \alpha + 4 \right ) \left ( 2 \alpha^2 \sigma \right )^{-1} \!  \int \! A^* \disp A \id V$ from \eqref{almostWaveAction} and using the identity
\beq
\int \! A^* \disp A \id V = \alpha \int \! \frac{f_0^2}{N^2} | A_z |^2 \id V - \int | \hnabla A |^2 \id V
\eeq
reveals the relationship
\beq
\action = \frac{\Ew}{\sigma} - \frac{\alpha + 4}{2 \alpha^2 \sigma} \int \! A^* \disp A \id V
\label{notQuiteClassical}
\eeq
between wave action $\action$ and energy $\Ew$.  The difference between action in the hydrostatic wave equation and $\Ew/\sigma$ depends on the fact that $\disp A \ne 0$.  Substituting equation \eqref{notQuiteClassical} into \eqref{almostAction} yields an equation for the evolution of wave energy, which is not conserved in the hydrostatic wave equation \eqref{internalTideEqnIntro}.  

Curiously, models that conserve wave action can be constructed with modifications to \eqref{cruderAsymptotics} that are similar to the modifications made in section \ref{remodeling}.  These action- and energy-conserving models lack either Galilean invariance or improved dispersion.  In some exploratory simulations, a model without improved dispersion fared worse and had a more limited regime of validity than equation \eqref{internalTideEqnIntro}.  Without Galilean invariance the model does not exactly describe Doppler shifting, though the consequences of such an inaccuracy have not been explored.  In section \ref{validationEnergy} we show that both $\action$ and $\Ew$ are nearly but not exactly conserved when a plane, vertical mode-one wave is distorted by two-dimensional turbulence.

\section{Validation}
\label{validation}

To build confidence in the validity of the hydrostatic wave equation \eqref{internalTideEqnIntro} we compare solutions to the linearized, hydrostatic Boussinesq equations and hydrostatic wave equation for a suite of initial value problems.  The initial value problems expose 20 vertical mode-one, horizontal plane waves with varying $\alpha$ to 3 two-dimensional turbulent flows with varying $\ep$.  Though this parameter study neglects the effects of vertical shear and buoyancy refraction, it nevertheless defines a region in $\alpha,\ep$ space where the model is accurate and provide a glimpse of how the hydrostatic wave equation fails as $\ep$ increases or $\alpha$ decreases.  

\subsection{The linearized hydrostatic Boussinesq equations and two-dimensional turbulence}

We linearize the hydrostatic Boussinesq equations around a two-dimensional mean flow 
\beq
\bU(x,y,t) = - \psi_y \bxh + \psi_x \byh
\label{streamfunctionDef}
\eeq
by substituting $\bu \mapsto \bU + \bu$ in \eqref{xmom}--\eqref{cont} and discarding terms quadratic in $\bu$ and $b$. These steps yield the set
\begin{align}
u_t + \bU \bcdot \bnabla u + \bu \bcdot \bnabla U - f_0 v + p_x &= 0 \com \label{xmomLin} \\
v_t + \bU \bcdot \bnabla v + \bu \bcdot \bnabla V + f_0 u + p_y &= 0 \com \label{ymomLin} \\
p_z &= b \com \label{zmomLin} \\
b_t + \bU \bcdot \bnabla b + w N^2 &= 0 \com \label{buoyLin} \\
u_x + v_y + w_z &= 0 \label{contLin} \per
\end{align}
Equations \eqref{xmomLin}--\eqref{contLin} describe the advection and refraction of waves by a two-dimensional flow with $\bU_{\! z} = \psi_z = 0$ and thus no buoyancy field.  The linearization neglects the complications of nonlinear wave dynamics and permits a two-dimensionalization of \eqref{xmomLin}--\eqref{contLin} by projection onto vertical modes.  Neither viscous dissipation in \eqref{xmomLin}--\eqref{zmomLin} nor diffusion in \eqref{buoyLin} is required to stabilize \eqref{xmomLin}--\eqref{contLin} for any of the solutions we report.  

The streamfunction $\psi$ in \eqref{internalTideEqnIntro} and \eqref{streamfunctionDef} obeys the two-dimensional vorticity equation with 4th-order hyperviscous dissipation,
\beq
\hlap \psi_t + \J \left ( \psi , \hlap \psi \right ) = - \nu_{\psi} \hlap^{\! 2} \! \left ( \hlap \psi \right ) \com
\label{vorticity}
\eeq
where $\nu_{\psi}$ is the hyperviscosity applied to $\hlap \psi$.  The solutions to \eqref{vorticity} we consider are relatively viscous and low resolution, but still exhibit characteristic features of geophysical and two-dimensional turbulence, such as persistent coherent vortices.

\subsection{The vertical mode decomposition}
\label{verticalModeProjection}

We restrict attention to waves with simple vertical structure by projecting \eqref{internalTideEqnIntro} and \eqref{xmomLin}--\eqref{contLin} onto the hydrostatic vertical modes $\mode_n(z)$ that solve the eigenproblem
\beq
\frac{f_0^2}{N^2} \mode_{nzz} + \kappa_n^2 \mode_n = 0 \com \qquad \text{with} \qquad \mode_n = 0 \quad \text{at} \quad z = -H, 0 \per
\label{modalEigenproblem}
\eeq
Note that the derivative $h_{nz}$ satisfies $\L h_{nz} = - \kappa_n^2 h_{nz}$.  The modal amplitudes of the independent variables $A, \bu, b, p$ are defined by their weighted projection onto $\mode_n$ or its derivative $\mode_{nz}$, with
\beq
\Phi_n \defn \int_{-H}^0 \Phi \, \mode_{nz} \id z  \qquad \text{for} \qquad \Phi = \left (A, u, v, p \right ) \com
\label{modezDef}
\eeq
and
\beq
b_n \defn \int_{-H}^0 b \, \mode_n \id z  \qquad \text{and} \qquad w_n \defn \int_{-H}^0 \frac{N^2 \kappa_n^2}{f_0^2} \, w \, \mode_n \id z \per
\eeq 
We assume $A, \bu, b$, and $p$ satisfy free-slip, rigid-lid homogeneous boundary conditions with $A_z = u_z = v_z = p_z = 0$ and $w= b = 0$ at $z = -H, 0$.
 
To project the hydrostatic wave equation \eqref{internalTideEqnIntro} onto the modes $\mode_{nz}$, we note that $\psi$ is two-dimensional and discard terms that depend on $\psi_z$, multiply by $\mode_{nz}$, integrate from $z=-H$ to $z=0$ and apply the definition of $A_n$ in \eqref{modezDef}.  We add 8th-order hyperviscosity to the result for numerical stability to obtain
\beq
\begin{split}
& \wave_n A_{nt} + \ii \alpha \sigma \disp_n A_n +  \J \left ( \psi, \wave_n A_n \right ) + \J \left ( A_n, \hlap \psi \right ) + \tfrac{\ii \sigma}{f_0} \hnabla \bcdot \left ( \hlap \psi \hnabla A_n \right ) \\
&  \qquad - \tfrac{2 \ii \sigma}{f_0^2} \left [ \J \left ( \psi_x, \ii \sigma A_{nx} - f_0 A_{ny} \right ) + \J \left ( \psi_y, \ii \sigma A_{ny} + f_0 A_{nx} \right )  \right ] = - \nu_A \hlap^{\! 4} \! \left ( \hlap A_n \right ) \com
\end{split}
\label{modeWiseWaveEqn}
\eeq
where $\nu_A$ is the hyperviscosity applied to $A_n$, and the mode-wise operators $\wave_n$ and $\disp_n$ are
\beq
\wave_n = \frac{\alpha}{2} \Big [ \hlap - \kappa_n^2 \left ( 4 + 3 \alpha \right ) \Big ] \qquad \text{and} \qquad \disp_n = \hlap + \alpha \kappa_n^2 \per
\eeq
Equation \eqref{modeWiseWaveEqn} describes the horizontal propagation of a mode-$n$ wave field with amplitude $A_n(x,y,t)$ through two-dimensional turbulence with streamfunction $\psi$.  The arbitrary stratification profile $N(z)$ enters \eqref{modeWiseWaveEqn} via the eigenvalue $\kappa_n^2$ determined by \eqref{modalEigenproblem}.

The linearized Boussinesq equations \eqref{xmomLin}--\eqref{contLin} are processed in similar fashion.  We project \eqref{xmomLin} and \eqref{ymomLin} onto $\mode_{nz}$, which yields
\begin{align}
u_{nt} - f_0 v_n + p_{nx} &= - \bU \bcdot \bnabla u_n - \bu_n \bcdot \bnabla U \com \label{modeWisexmom} \\
v_{nt} + f_0 u_n + p_{ny} &= - \bU \bcdot \bnabla v_n - \bu_n \bcdot \bnabla V \per \label{modeWiseymom}
\end{align}
We next combine \eqref{zmomLin}--\eqref{contLin} by projecting \eqref{contLin} onto $\mode_{nz}$, integrating by parts once, and using \eqref{modalEigenproblem} to yield $w_n = - u_{nx} - v_{ny}$.  Then using $p_z = b$ to combine \eqref{zmomLin} and \eqref{buoyLin}, projecting the result onto $\mode_n$, integrating by parts and substituting $w_n = - u_{nx} - v_{ny}$ leads to
\beq
p_{nt} + \left ( \tfrac{f_0}{\kappa_n} \right )^{\! 2} \left ( u_{nx} + v_{ny} \right ) = - \bU \bcdot \bnabla p_n \per
\label{modeWisebuoy}
\eeq
The three equations \eqref{modeWisexmom}--\eqref{modeWisebuoy} describe the evolution of hydrostatic, vertical mode-$n$ waves in a two-dimensional flow $\bU = U \bxh + V \byh$ with $\bU_{\! z} = 0$.  The parameter $f_0 / \kappa_n$ is the phase speed of a linear wave with mode-$n$ vertical structure.  

\subsection{Initial value problems and numerical methods}
\label{setup}

We solve \eqref{vorticity} simultaneously with \eqref{modeWiseWaveEqn} and \eqref{modeWisexmom}--\eqref{modeWisebuoy} for a series of initial value problems that place a horizontal plane wave with the vertical structure of a single vertical mode into mature two-dimensional turbulence in a doubly periodic domain.  The periodic physical domain is square with dimension $L = 1600 \, \mathrm{km}$, which fits 16 wavelengths of a plane wave with dimensional wavenumber $\kw = \pi / 50 \, \mathrm{km}^{-1}$.  In varying $\alpha$ from 0.1 to 2, we fix the domain size $L$, wavenumber $\kw$, initial turbulent field $\psi$, and inertial frequency $f_0 = 10^{-4} \, \mathrm{s^{-1}}$ while co-varying $\kappa_n = \kw / \sqrt{\alpha}$ and $\sigma = f_0 \sqrt{1 + \alpha}$ with $\alpha$.  

The initial condition for $A_n$, 
\beq
A_n \, \big |_{t=0} = \ee^{\ii \kw x} a \com 
\label{initialConditionA}
\eeq
excites a rightward propagating horizontal plane wave.  In \eqref{initialConditionA} $a$ is the constant initial magnitude of $A_n$ and $\kw = \pi / 50 \, \mathrm{km}^{-1}$ is the wave field's initial wavenumber.  The linearized nature of both \eqref{modeWiseWaveEqn} and \eqref{modeWisexmom}--\eqref{modeWisebuoy} means the initial magnitude of the wave field is arbitrary; we choose $a = \alpha f_0 / 2 \kw \sqrt{\alpha + 2}$ to produce an initial maximum speed $\max \big ( \sqrt{u_n^2+v_n^2} \big ) = 1 \, \mathrm{m \, s^{-1}}$.

The initial conditions for $p_n$, $u_n$, and $v_n$ in \eqref{modeWisexmom}--\eqref{modeWisebuoy} are
\beq 
\big [ \, p_n, u_n, v_n \big ]_{t=0} = \frac{2 a}{\alpha f_0^2} \Big [ \, \alpha f_0^3 \cos \left ( \kw x \right ), \,  \kw \sigma \cos \left ( \kw x \right ), \, \kw f_0 \sin \left ( \kw x \right ) \Big ] 
\label{initialConditionV}
\eeq
corresponding to the same progressive plane wave in \eqref{initialConditionA} with the mode-$n$ pressure field $p_n = 2 a f_0 \cos \big ( \kw x - \sigma t \big )$ at $t=0$.

We generate three turbulent initial conditions for $\psi$ by integrating \eqref{vorticity} from the random state 
\beq
\hat \psi \, \big |_{t=-T} =  \frac{\Psi \ee^{\ii \theta} \sqrt{k^2 + \ell^2}}{\big ( 1 + k_c^{-1}\sqrt{k^2+\ell^2}  \, \big )^{8}} \com
\label{initialPsi}
\eeq
for a preliminary interval of length $T$ up to $t=0$.  In \eqref{initialPsi} $\hat \psi(k,\ell,t)$ is the two-dimensional Fourier transform of $\psi(x,y,t)$ and $\theta(k,\ell)$ is the random initial phase of wavenumber $k, \ell$.  We choose the dimensional value $k_c = 64 \times 2 \pi / L$ in \eqref{initialPsi} so that the energy spectra $\left ( k^2 + \ell^2 \right ) | \hat \psi |^2$ is initially concentrated around non-dimensional wavenumber 64.  Three magnitudes $\Psi$ in \eqref{initialPsi} are chosen so the random state in \eqref{initialPsi} has the root-mean-squared Rossby numbers $\text{r.m.s.} \! \left ( \hlap \psi / f_0 \right ) = ( 0.07, 0.1, 0.2 )$.  The resulting random states are then integrated for the preliminary intervals $T = ( 600, 400, 200 ) \times 2 \pi / f_0 \, \, \mathrm{s}$, respectively, to produce turbulent initial conditions $\psi(t=0)$ with the the properties $\text{max} \left ( \hlap \psi / f_0 \right ) \approx ( 0.033, 0.064, 0.14 )$ and $\text{max} \big ( | \bnabla \psi | \kw / f_0 \big ) \approx ( 0.039, 0.060, 0.12 )$.  The parameters and intervals used for the preliminary integrations are tuned so that $\max \left ( \hlap \psi / f_0 \right )$ and $\max \big ( | \bnabla \psi | \kw / f_0 \big )$ are similar for each of the initial turbulent states, which implies that all terms in \eqref{modeWiseWaveEqn} are of comparable importance.  Hereafter we use $\max \left ( \hlap \psi / f_0 \right )  \approx ( 0.033, 0.064, 0.14 )$ as reference values for $\ep$. 

Equations \eqref{vorticity}, \eqref{modeWiseWaveEqn} and \eqref{modeWisexmom}--\eqref{modeWisebuoy} are solved on a square doubly-periodic domain using a dealiased pseudospectral method with $256^2$ Fourier modes in $x$ and $y$.  The ETDRK4 scheme described by \cite{cox2002exponential}, \cite{KassamTrefethen}, and \cite{grooms2011linearly} is used to numerically integrate equations  \eqref{vorticity}  and \eqref{modeWiseWaveEqn} in time, while a 4th-order Runge-Kutta scheme is used to integrate the modal hydrostatic Boussinesq equations \eqref{modeWisexmom}--\eqref{modeWisebuoy}.  We use the hyperviscosities $\nu_{\psi} = 3 \times 10^{8} \, \mathrm{m^4 \, s^{-1}}$ in \eqref{vorticity} and $\nu_A = 10^{24} \, \mathrm{m^8 \, s^{-1}}$ in \eqref{modeWiseWaveEqn}.  Due to hyperdissipation the three turbulent fields lose 1-3\% of their energy at $t=0$ over the few hundred wave periods that we consider.

\subsection{Wave field evolution with $\alpha = 1$ and $\ep \approx 0.14$} 

\begin{figure}
\centering
\includegraphics[width = 1\textwidth]{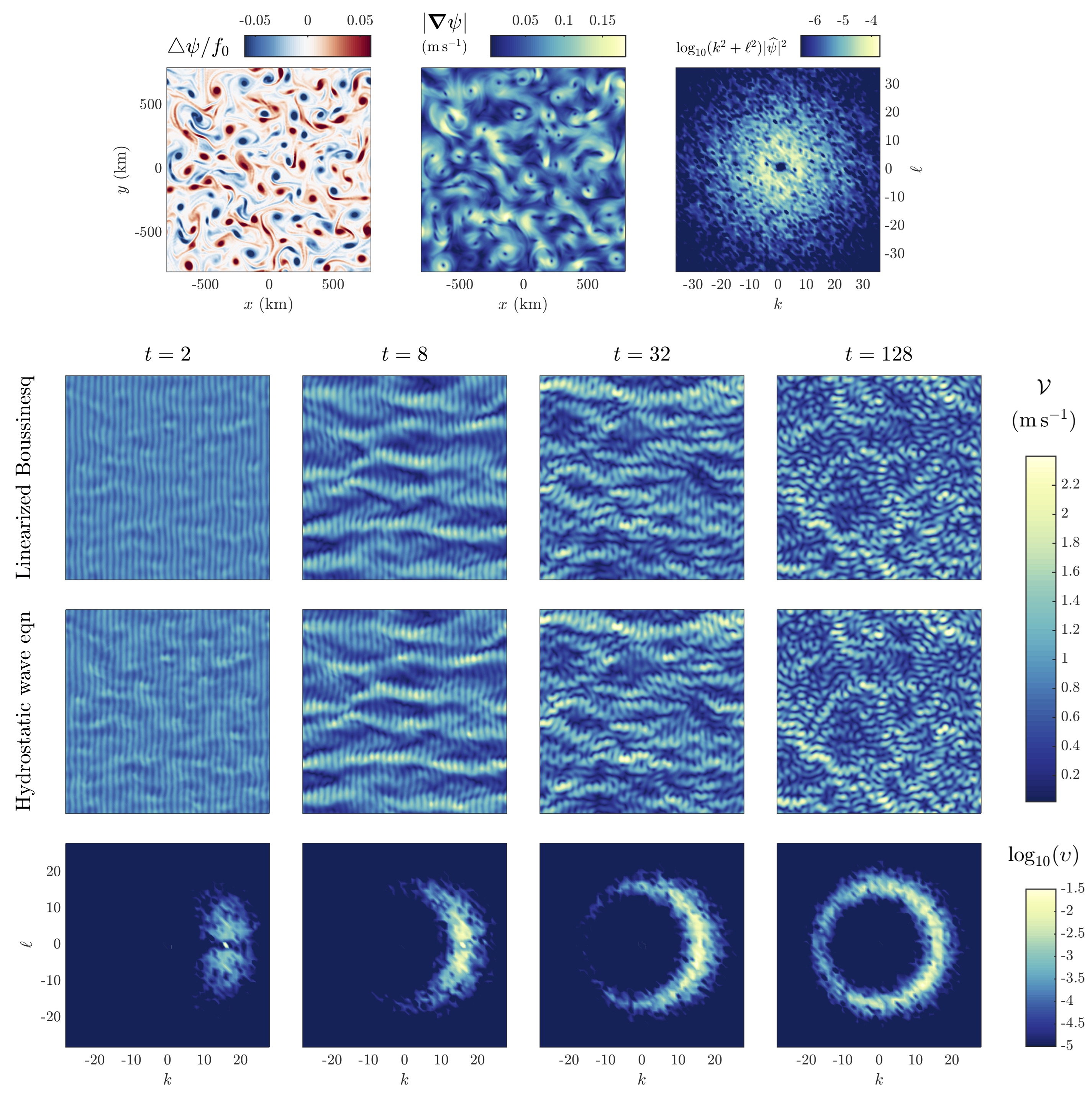}
\caption{Scattering of a plane wave with frequency $\sigma = f_0 \sqrt{2}$ and thus $\alpha = 1$ by two-dimensional turbulence with maximum vorticity $\text{max}(\hlap \psi / f_0) \approx 0.14$ in the linearized Boussinesq equations and the hydrostatic wave equation.  Parameters and initial conditions are given in section \ref{setup}.  The top 3 panels from left to right show the initial turbulent vorticity, speed, and energy spectra.  The bottom 12 panels show wave field evolution in four snapshots: the first row shows speed $\speed_B$ in the linearized hydrostatic Boussinesq system \eqref{modeWisexmom}--\eqref{modeWisebuoy}; the second row shows speed $\speed_A$ in the hydrostatic wave equation \eqref{modeWiseWaveEqn}; and the third row shows the logarithm of the spectral measure $\E_A$ from the hydrostatic wave equation.  $\speed$ and $\E$ are defined in \eqref{speedDef} and \eqref{specDef}.} 
\label{evolution}
\end{figure}

The initial turbulent field and the evolution of $A_n$ in the hydrostatic wave equation and $u_n, v_n$, and $p_n$ in the linearized Boussinesq equations are shown in figure \ref{evolution} for a case with wave Burger number $\alpha = 1$ and Rossby number $\ep \approx \max \left ( \triangle \psi / f_0 \right ) \approx 0.14$.  The top row of figure \ref{evolution} shows the initial normalized turbulent vorticity $\hlap \psi / f_0$, speed $| \bnabla \psi |$, and energy spectra $(k^2 + \ell^2 ) | \hat \psi |^2$ from left to right.  Turbulent vorticity is concentrated in coherent vortices and turbulent energy in non-dimensional wavenumbers less than $\sqrt{k^2 + \ell^2} \approx 8$.  As a result, wave field spectral components experience a gradual diffusion to nearby wavenumbers rather than the sharper reflection that a smaller-scale turbulent field would incur.  Hereafter in figures and text the wavenumbers $k$ and $\ell$ denote non-dimensional Fourier wavenumbers normalized by $2 \pi / L$. 

The bottom three rows portray the turbulent scattering of the initially planar wave field in four snapshots at $t = 2$, 8, 32, and 128 wave periods.  The second and third rows of figure \ref{evolution} show snapshots of mode-wise wave speed,
\beq
\speed(x,y,t) \defn \sqrt{u_n^2+v_n^2} \com
\label{speedDef}
\eeq
which is diagnosed from the hydrostatic wave equation solution using the leading-order relations in \eqref{leadingOrderVelocity}.  We use subscripts to differentiate between models, so that $\speed_B$ is diagnosed from the linearized hydrostatic Boussinesq system \eqref{modeWisexmom}--\eqref{modeWisebuoy}, and $\speed_A$ from the hydrostatic wave equation \eqref{modeWiseWaveEqn}.  The bottom row shows snapshots of the normalized wave potential energy spectra
\beq
\E(k,\ell,t) \defn \frac{ | \hat A_n |^2}{\int | \hat A_n |^2 \id k \id \ell}
\label{specDef}
\eeq
from the hydrostatic wave equation \eqref{modeWiseWaveEqn}.

The snapshots of speed $\speed$ and spectra $\E$ reveal how wave scattering by turbulence leads both to an isotropization of wave energy around the circle $k^2 + \ell^2 = \kw^2$ as well as smearing of the energy spectrum to wavenumber magnitudes higher and lower than $\kw$.  The smearing of energy around $\kw$ indicates the importance of near-resonant interactions between waves and turbulence.  At $t=2$ most of the energy is concentrated at $k = \kw$.  By $t = 8$ the initial stages of isotropization are underway, attended by a focusing and concentration of wave energy in strips parallel to the original direction of wave propagation.  Focusing is generic in the scattering of parallel incident waves, especially in the geometrics optics limit \citep{white1998chance, nye1999natural}.  As the isotropization proceeds, random focusing gives way to almost-isotropic disorder by $t = 128$. 

The agreement between the two models is impressive: excellent correspondence both in the spatial structure and quantitative amplitude of wave field energy persists to $t = 128$ wave periods.  Interestingly, the most obvious differences in wave speed are at the earliest time $t = 2$ wave periods.  The pointwise comparison of wave speed over hundreds of wave periods is a severe test of the asymptotic model, and correspondences between wave field spectra and statistics diagnosed from the two models for the same parameters are closer still.  We find that for the parameters explored here, such striking validity holds approximately when $\ep / \alpha < 0.2$.  For larger values of $\ep/\alpha$ nonlinear advection and refraction overcome the effects of dispersion, which consequently leads to non-small $\D A$, disrupts the assumed ordering of terms in the wave operator equation \eqref{waveOperatorFormTide}, and invalidates the assumptions used to derive \eqref{internalTideEqnIntro}.

\subsection{Physical-space and statistical comparisons across $\alpha,\ep$ parameter space}

We next explore the $\alpha, \ep$  parameter space with 60 simulations of both the hydrostatic wave equation \eqref{modeWiseWaveEqn} and linearized Boussinesq system \eqref{modeWisexmom}--\eqref{modeWisebuoy}.  The 60 cases correspond to 20 equispaced values of $\alpha$ between $\alpha = 0.1$ and $\alpha = 2$ for each of the 3 turbulent vorticity fields with $\ep \approx 0.033$, 0.064, and 0.14.  We compare physical fields and spectra of the two models before using a bulk measure of physical space error in solutions to the hydrostatic wave equation to compare the results in aggregate.

\begin{figure}
\centering
\includegraphics[width = 1\textwidth]{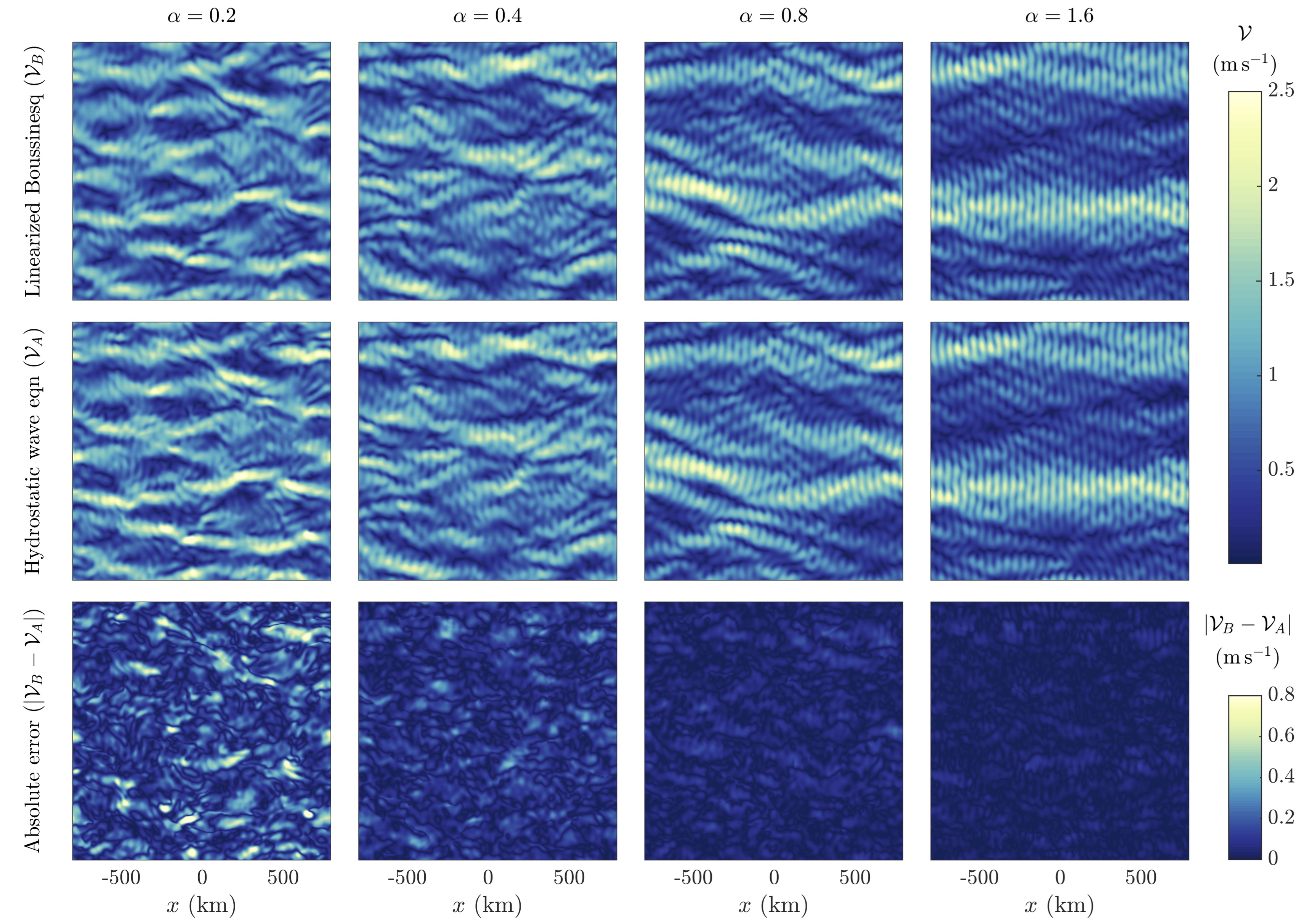}
\caption{A qualitative physical-space comparison between snapshots of wave speed from the linearized Boussinesq equations and the hydrostatic wave equation for four initial value problems with wave Burger numbers $\alpha = 0.2$, 0.4, 0.8, and 1.6.  The snapshots are taken at $t = 10 \alpha$ wave periods.  The initial value problems expose an initially planar wave field to a two-dimensional turbulent flow with $\ep \approx \max \left ( \hlap \psi / f_0 \right ) \approx 0.064$ and are described in section \ref{setup}.  The top and middle rows show wave speed $\speed_B$ from the linearized Boussinesq system and $\speed_A$ from the hydrostatic wave equation, respectively, and the bottom row shows the absolute error $|\speed_B-\speed_A|$.}
\label{qualitativeComparison}
\end{figure}

Figure \ref{qualitativeComparison} compares snapshots of wave speed $\speed$ from four linearized Boussinesq and hydrostatic wave equation solutions with $\alpha = 0.2$, 0.4, 0.8, and 0.16 and $\ep \approx 0.064$ at $t = 10\alpha$ wave periods.  The top row of figure \ref{qualitativeComparison} shows wave speed $\speed_{B}$ defined in \eqref{speedDef} from the linearized Boussinesq equations, the middle row shows $\speed_{A}$ from the hydrostatic wave equation, and the bottom row shows the absolute error $|\speed_{B} - \speed_{A}|$ between the two.  The results show clearly that for fixed $\ep$ the error decreases when $\alpha$ increases; when $\alpha = 1.6$ and $\ep \approx 0.064$ the pointwise error in wave speed after $t = 160$ wave periods is almost everywhere less than 10\% of its initial value.  Despite the relatively large errors when $\alpha = 0.2$, the spatial structure of $\speed$ is broadly similar between both models. 

\begin{figure}
\centering
\includegraphics[width = 1\textwidth]{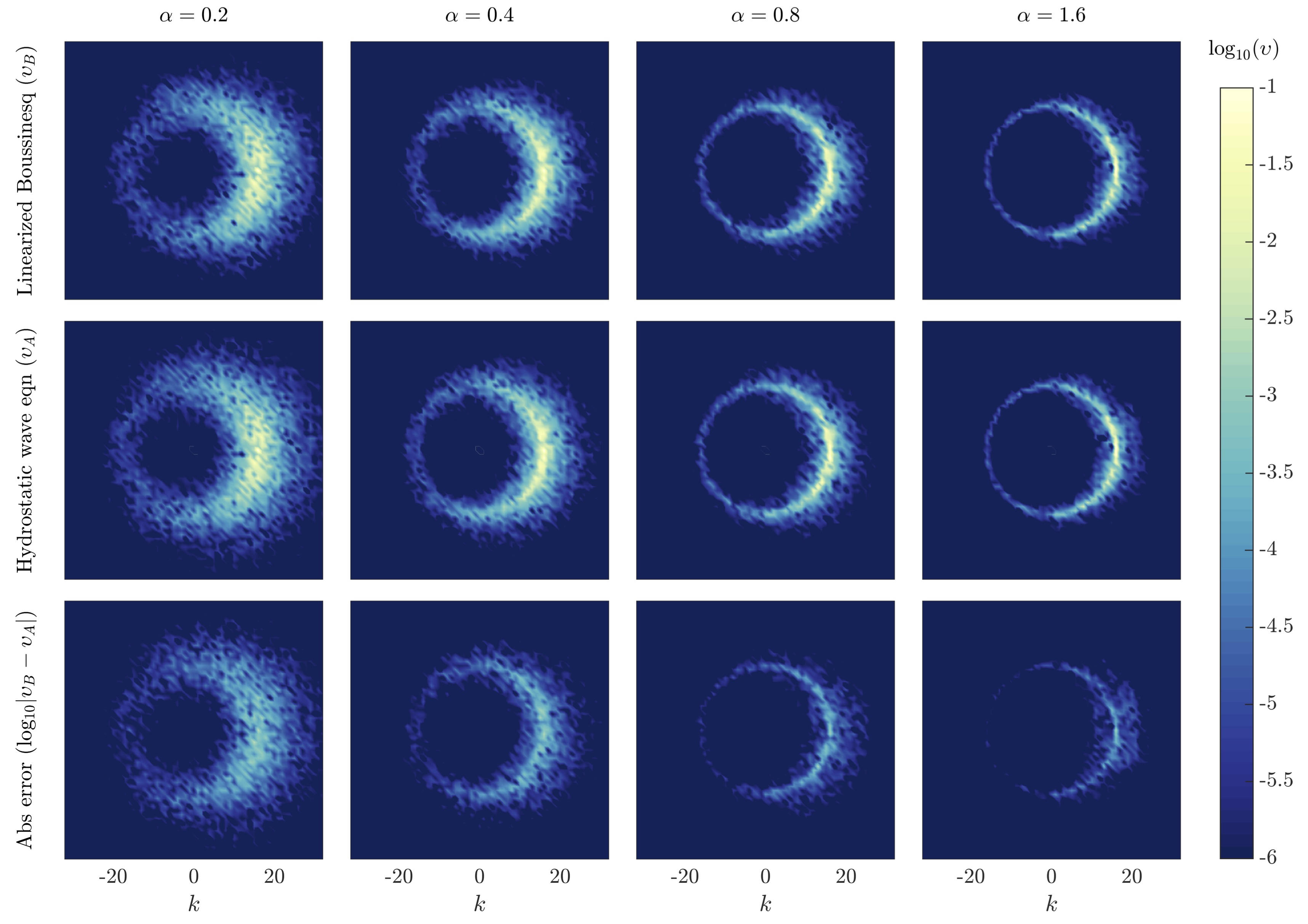}
\caption{Comparison of normalized potential energy spectral amplitudes $\E$ defined in \eqref{specDef} for the same simulations considered in figure \ref{qualitativeComparison}.  The top row shows $\E_B$ from the linearized Boussinesq equations, the middle row shows $\E_A$ from the hydrostatic wave equation and the bottom row shows the absolute difference $| \E_B - \E_A|$, all scaled logarithmically.}
\label{qualitativeSpectralComparison}
\end{figure}

The pointwise comparison of wave speed $\speed$ is a strict test of model accuracy.  We move toward less stringent statistical comparisons with figure \ref{qualitativeSpectralComparison}, which replicates the form of figure \ref{qualitativeComparison} for snapshots of the normalized spectral amplitudes $\E$ defined in \eqref{specDef} in terms of $\hat A_n$.  To estimate $A_n$ from the linearized Boussinesq solution, we observe that the definition of $A$ in terms of $p$ in \eqref{leadingOrderPressure} implies that
\beq
p_{nt} = - \ii \sigma \left ( \ee^{ - \ii \sigma t} A_n - \ee^{\ii \sigma t} A^*_n \right ) + \ee^{- \ii \sigma t} A_{nt} + \ee^{\ii \sigma t} A^*_{nt} \per
\label{pressureTimeDerivative}
\eeq
Due to the slow variation of $A_n$, which implies that $A_{nt} / \sigma A_{n} \sim \ep \ll 1$, the parenthetical terms in \eqref{pressureTimeDerivative} are both $O(\ep^{-1})$ larger than the two rightmost terms.  This implies the approximate formula for $A_n$
\beq
A_n \approx \frac{\ee^{\ii \sigma t}}{2 f_0} \left ( p_n + \ii \sigma^{-1} p_{nt} \right ) \com
\label{approximateA}
\eeq
in terms of the linearized Boussinesq variable $p_n$ and $p_{nt}$.  The Fourier transform of \eqref{approximateA} provides an estimate of $\hat A_n$ from $\hat p_n$.

The top two rows of figure \ref{qualitativeSpectralComparison} show $\E_B$ from the linearized Boussinesq system and $\E_A$ hydrostatic wave equation, scaled logarithmically.  The spectral amplitudes $\E$ for each model are remarkably similar.  The bottom row of figure \ref{qualitativeSpectralComparison} shows the absolute difference $| \E_B - \E_A |$ between the top two rows.  Spectral errors are small and decrease with increasing $\alpha$ for fixed $\ep$.

We next isolate specific modes of model failure by moving from the non-dimensional Cartesian spectral coordinates $k, \ell$ into the polar spectral coordinates $\varkappa,\theta$ defined so that $k = \varkappa \cos \theta$ and $\ell = \varkappa \sin \theta$.  We define the spectral integral measure $\rE$ as
\beq
\rE(\varkappa,t) \defn \int_{0}^{2 \pi} | \hat A_n |^2 \, \varkappa \id \theta \per
\label{polarSpectraDef}
\eeq
$\rE$ is similar to the one-dimensional energy spectra used to analyze two-dimensional turbulence, and the integral $\int \rE \id \varkappa$ is proportional to total wave field potential energy.  $\rE$ reveals the radial distribution of $| \hat A_n |^2$ and thus measures the spatial scales in $\hat A_n$ regardless of the direction of propagation of the mode $k,\ell$.    To calculate $\Upsilon$ numerically we interpolate $\hat A_n$ known at discrete $k,\ell$ values onto a $1024 \times 256$ grid in $\varkappa,\theta $ and integrate  $| \hat A_n |^2$ over $\theta$.  

\begin{figure}
\centering
\includegraphics[width = 1\textwidth]{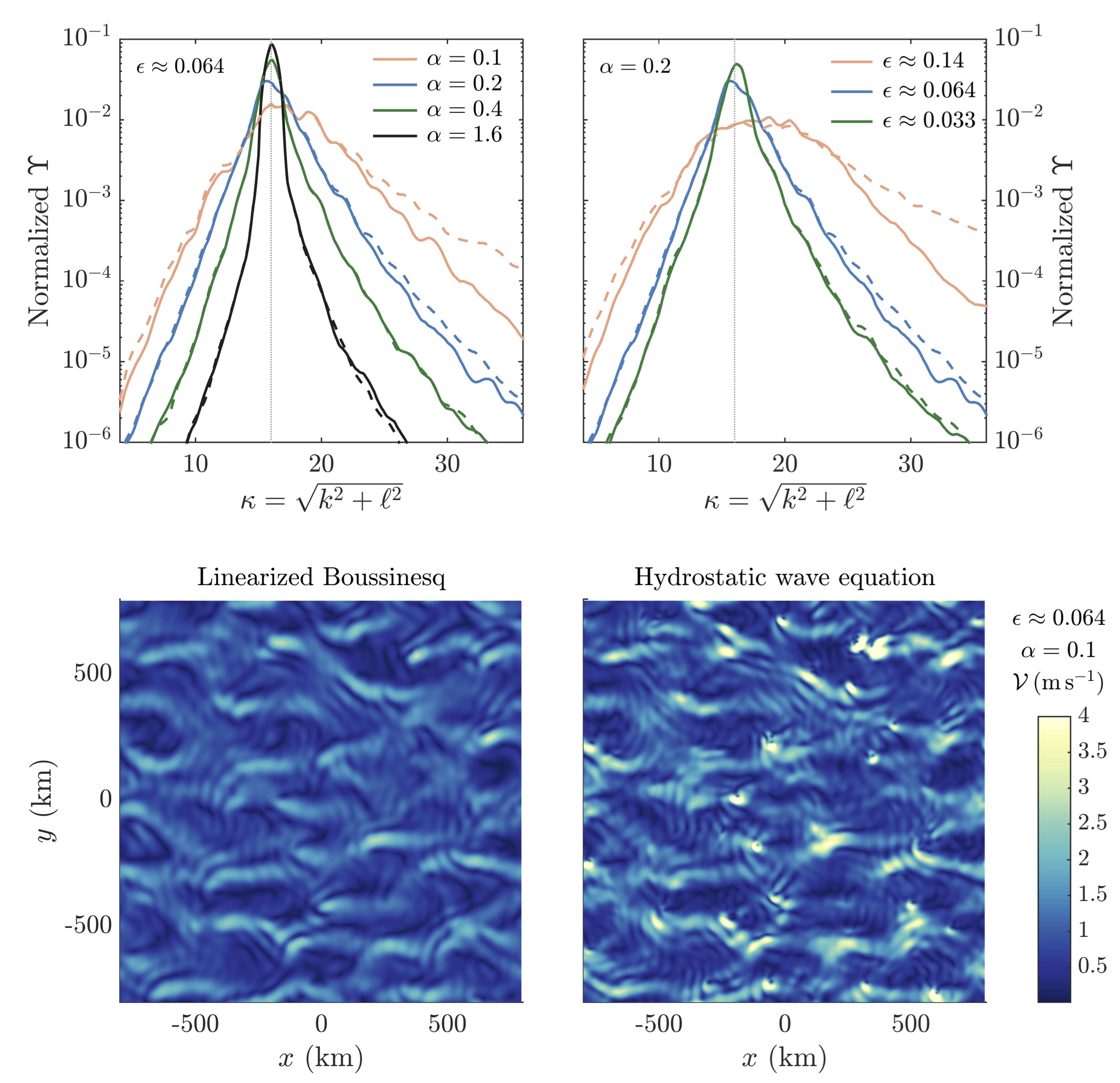}
\caption{Comparison of the polar-integrated spectral measure $\rE(\varkappa)$ defined in \eqref{polarSpectraDef} in the linearized Boussinesq equations (solid lines) and hydrostatic wave equation (dashed lines), both normalized by $\int \rE \id \varkappa$ from the linear Boussinesq result.  The top left panel compares four solutions with $\ep \approx 0.064$ with varying $\alpha$ while the top right panel compares three solutions with $\alpha = 0.2$ and varying $\ep$.  A dotted line indicates the wave field's initial wavenumber $\varkappa = 16$.  The bottom panels show wave speed $\speed$ from the two models for the  case $\ep = 0.064$ and $\alpha = 0.1$ to illustrate the spuriously intense small-scale features that develop in the hydrostatic wave equation solution when $\alpha/\ep$ is small.  All snapshots are taken at $t \approx 13\alpha/\ep$ wave periods.}
\label{polarSpectra}
\end{figure}

Figure \ref{polarSpectra} shows snapshots of $\rE$ at $t \approx 13\alpha/\ep$ wave periods for six cases with varying $\alpha$ and $\ep$: the top left panel holds $\ep \approx 0.064$ constant and varies $\alpha$, while the top right panel holds $\alpha = 0.2$ constant and varies $\ep$.  In both panels $\rE$ is normalized by $\int \rE \id \varkappa$ from the linearized Boussinesq solution.  The bottom left and right panels compare snapshots of $\speed_B$ and $\speed_A$ for the case $\ep = 0.064$ and $\alpha = 0.1$.  The $\rE$ comparisons reveal aspects both of wave-flow interaction and the errors that develop in the hydrostatic wave equation for small $\alpha/\ep$: first, because exactly `resonant' wave-flow interactions only redistribute energy among wave modes with $\varkappa = 16$, the width of $\rE$ associated with energy at off-dispersion wavenumbers around $\varkappa = 16$ is due explicitly to near-resonant dynamics.  Second, all curves are asymmetric about the central wavenumber $\varkappa = 16$, showing that these near-resonant interactions preferentially shift energy to higher wavenumbers.  Third, the most severe errors in $\rE$ in the hydrostatic wave equation are associated with an over-prediction of wave energy at very high wavenumbers.  The worst-case comparison in the bottom panels of figure~\ref{polarSpectra} shows how these errors manifest as regions of spuriously intense small-scale wave activity.

\begin{figure}
\centering
\includegraphics[width = 1\textwidth]{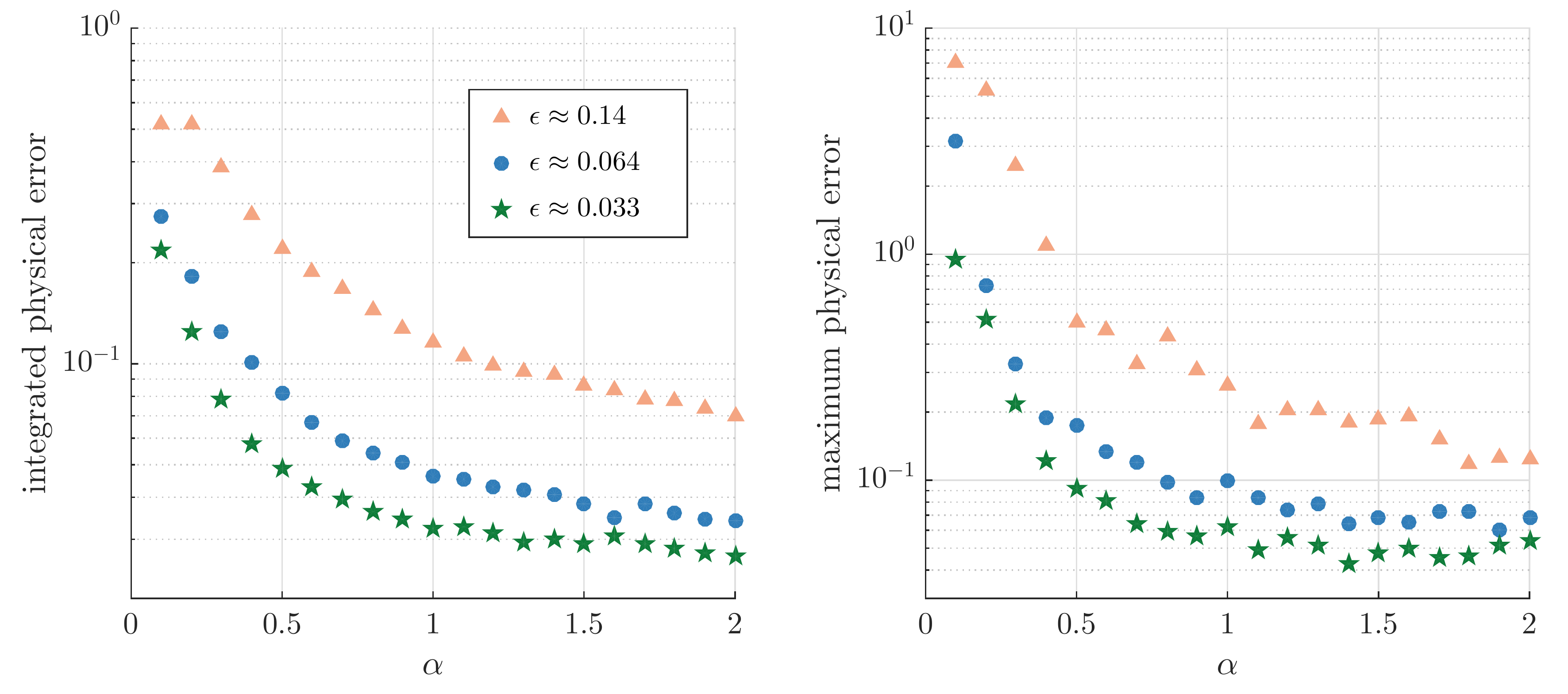}
\caption{Integrated and maximum point-wise error in 60 solutions to the hydrostatic wave equation corresponding to 3 values of $\ep \approx \max \left ( \triangle \psi / f_0 \right )$ and 20 values of $\alpha$.  For every solution the error is computed at $t \approx 6.5 \alpha / \ep$ wave periods.  The integrated error is defined in \eqref{integratedError} and maximum error is the maximum point-wise error in speed defined in \eqref{maximumError}.}
\label{synthesis}
\end{figure}

We finally aggregate all solutions by introducing two bulk metrics: the `integrated error' and `maximum error'.  Integrated error measures the total sum of errors in snapshots of wave speed and is defined by
\beq
\text{integrated error} \defn \frac{ \int \big | \speed_B - \speed_A \big | \id x \id y }{\int  \speed_B \id x \id y} \per
\label{integratedError}
\eeq
The maximum error defined by
\beq
\text{maximum error} \defn \frac{ \max | \speed_B - \speed_A |}{\max \left (  \speed_B \right ) }
\label{maximumError}
\eeq
isolates the worst-case relative errors in wave speed at particular locations and times.  Figure \ref{synthesis} shows snapshots of integrated error and maximum error for all 60 initial value problems as a function of $\alpha$.  The snapshots are taken at the approximate time $t \approx 6.5 \alpha / \ep$ wave periods.  All errors decrease both as $\ep$ decreases and as $\alpha$ increases.  The maximum error in the physical space solution is less than 10\% when $\ep \le 0.064$ and $\alpha \ge 0.8$, but is never less than $10\%$ when $\ep \approx 0.14$ for the range of $\alpha$ and time-snapshots considered.  Maximum errors increases sharply for small $\alpha$ and are more than 50\% when $\alpha \le 0.2$ for all $\ep$.

\subsection{The evolution of wave energy and action}
\label{validationEnergy}

\begin{figure}
\centering
\includegraphics[width = 1\textwidth]{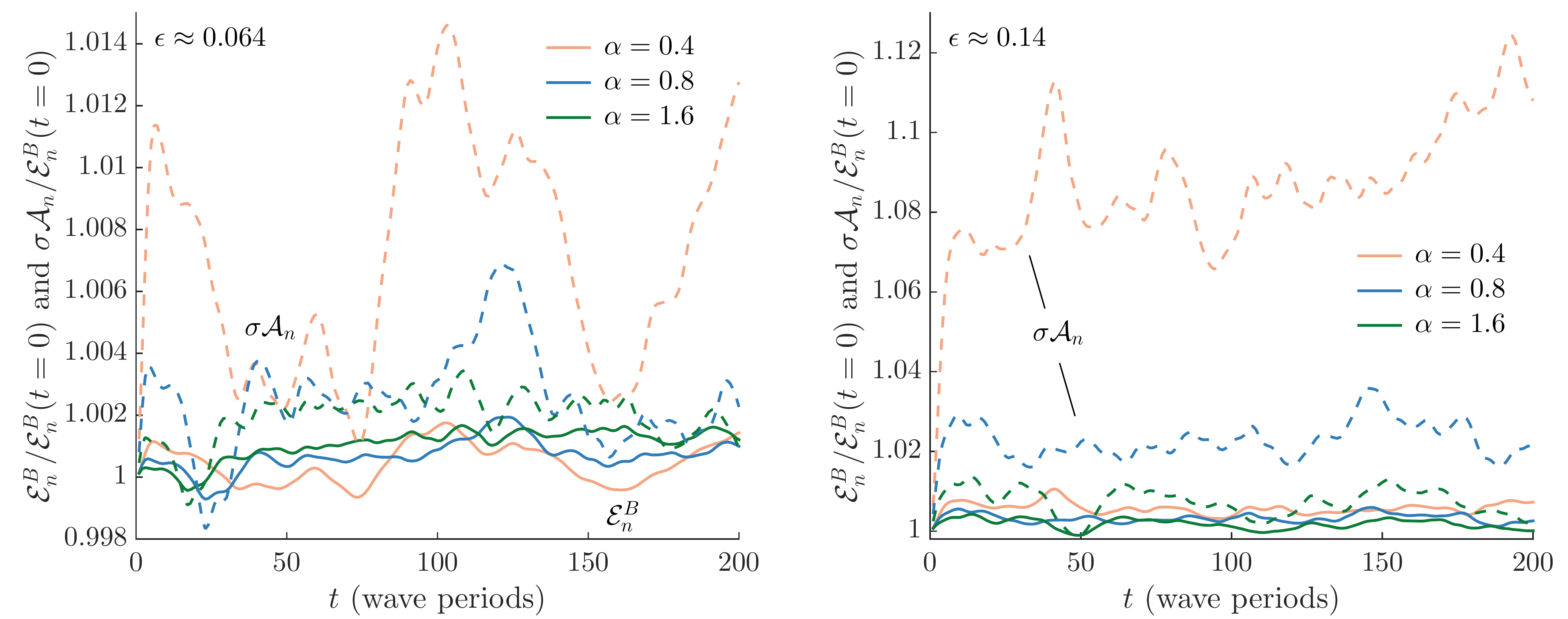}
\caption{Comparison of $\sigma \action_n$ (dashed lines) and $\mathcal{E}^B_n$ (solid lines), both normalized by the initial wave energy $\mathcal{E}^B_n(t=0)$.  The left panel shows three cases with $\alpha = 0.4$, 0.8, and 1.6 with $\epsilon \approx 0.064$, and the right panel shows three cases with the same $\alpha$ and $\ep \approx 0.14$.  Both $\mathcal{A}$ and $\mathcal{E}^B$ are conserved to within a few percent in all cases except $\alpha=0.4$ and $\ep \approx 0.14$.  Note that the panels have different $y$-axes.}
\label{energy}
\end{figure}

We turn at last to the transfer of energy between waves and turbulence.  We use the evolution of wave action $\action$ defined in \eqref{almostAction} to diagnose energy transfers in the hydrostatic wave equation.  The mode-wise version of $\action$ is
\beq
\mathcal{A}_n = \frac{1}{2 \alpha \sigma} \int | \hnabla A_n |^2 + \left ( 4 + 3 \alpha \right ) \kappa_n^2 | A_n |^2 \id x \id y \per
\eeq
An equation for the evolution of wave energy density in the linearized Boussinesq equations follows from the combination $u_n \eqref{modeWisexmom} + v_n \eqref{modeWiseymom} + \left ( \kappa_n / f_0 \right )^2 \! p_n \eqref{modeWisebuoy}$, which produces
\beq
\pe_{nt} + \bnabla \bcdot \left ( \bu_n p_n + \bU \pe_n \right ) = - u_n \bu_n \bcdot \bnabla U - v_n \bu_n \bcdot \bnabla V \com
\label{perturbationEnergyDensityEvolution}
\eeq
where wave energy density is defined
\beq
\pe _n\defn \half u_n^2 + \half v_n^2 + \tfrac{\kappa_n^2}{2 f_0^2} p_n^2 \per
\eeq
The superscript `$B$' stands for `Boussinesq'.  The total mode-wise wave energy $\pE_n \defn \int \pe_n \id V$, which is not conserved in \eqref{modeWisexmom}--\eqref{modeWisebuoy} due to the non-zero right side of \eqref{perturbationEnergyDensityEvolution}, is therefore
\beq
\pE_n = \half \int u_n^2 + v_n^2 + \tfrac{\kappa_n^2}{f_0^2} p_n^2 \id x \id y \per
\label{integratedPerturbationEnergy}
\eeq
We compare the evolution of $\sigma \action_n$ and $\pE_n$, which are initially equal for the initial conditions in \eqref{initialConditionA}--\eqref{initialConditionV} because $\disp_n A_n \, |_{t=0} = 0$.  The product $\sigma \action_n$ and wave energy in the hydrostatic wave equation are closely related by the identity in \eqref{notQuiteClassical}.  

Our comparison is summarized in figure~\ref{energy}, which shows the evolution of $\sigma \action_n$ and $\pE_n$ both normalized by total initial wave energy $\pE_n(t=0)$ for three values of $\alpha = 0.4, 0.8, 1.6$.  The right panel corresponds to the case $\ep \approx 0.064$ and the left panel to $\ep \approx 0.14$.  Even in the most nonlinear case with $\ep \approx 0.14$ the energy of the linearized Boussinesq solution remains within 1\% of its initial value: in other words, there is almost no transfer of energy between waves and flow in these non-near-inertial cases.  The comparison shows also that the mode-wise wave action $\action_n$ is nearly conserved when $\ep / \alpha$ is small.  The $\sim10\%$ change in $\A_n$ at $\ep \approx 0.14$ and $\alpha = 0.4$ betrays the strong increases in $\action_n$ that manifest when $\ep / \alpha$ approaches unity.

 Previous discussions of energy transfer between waves and quasi-geostrophic flow  \citep{buhler2005wave,polzin2010mesoscale} did not prepare us for the discovery that there is almost no transfer of energy in the linearized Boussinesq system.  A crucial feature of figure~\ref{evolution} is that wave energy does not cascade to small length scales: the main impact of turbulent distortion is the formation of `wave dislocations' \citep{NyeBerry}. 
 
In summary, both the hydrostatic wave equation and the linearized Boussinesq system exhibit weak energy transfers between waves and turbulence, and the small transfers in the hydrostatic wave equation are systematically larger than those in linearized Boussinesq system. In the least-accurate case in figure \ref{energy} where $(\alpha,\ep) = (0.4,0.14)$,  the hydrostatic wave equation has transfers on the order $7$--$11\%$, while the Boussinesq transfers are always less than $1\%$.  We speculate that increasing $\epsilon$ will result in larger transfers, but characterization of these transfers lies beyond our present scope.

\subsection{Summary of section \ref{validation}}

The hydrostatic wave equation provides an accurate approximation of linearized Boussinesq dynamics when $\ep/\alpha$ is small, or when the wave frequency is sufficiently far from inertial and the quasi-geostrophic flow is weak enough in combination.  For example, here the maximum error is everywhere less than 10\% when $\ep \le 0.064$ and $\alpha \ge 0.8$.  Conversely, great care must be taken in using \eqref{internalTideEqnIntro} when the wave field approaches near-inertial: when $\alpha < 0.5$ and $\sigma < 1.22 f_0$, maximum error in the hydrostatic wave equation less than 10\% only when the mean flow is very weak and $\ep \le 0.033$. Failures of the hydrostatic wave equation are systematically associated with too-large transfers of wave energy to high wavenumbers and the subsequent development of spuriously-small spatial scales in the wave field.  Yet even when the hydrostatic wave equation does not well-predict wave field spatial structure it may provide a decent approximation of wave field statistics such as the spectral distribution of wave energy.  Finally, for the cases we consider waves and turbulence exchange only small amounts of energy.

\section{Discussion}
\label{discussion}

This paper introduces the `hydrostatic wave equation': a new reduced model for the propagation of three-dimensional hydrostatic internal waves through quasi-geostrophic flow.  The hydrostatic wave equation detailed in section \ref{summary} and exhibited in \eqref{internalTideEqnIntro} is appropriate for describing the propagation of non-inertial  internal tides of arbitrary scale through the inhomogeneous ocean.  The primary virtue of the hydrostatic wave equation is the filtering of fast wave oscillations.  This phase averaging  isolates wave advection and refraction on the slow time scales of quasi-geostrophic flow evolution and permits the use of relatively large time-steps in numerical solutions.  Time-filtering thus facilitates both computations and theoretical analysis, such as an estimate of internal tide scattering rates similar to that applied to Young and Ben Jelloul's near-inertial equation by \citet{PhysRevFluids.1.033701}.  The costs of filtering are the errors that emerge when the mean flow is too strong.  

The most important ingredient in the derivation of \eqref{internalTideEqnIntro} is the reconstitution of the leading-order dispersion constraint with the first-order effects of wave advection and refraction by quasi-geostrophic flow.  Because of reconstitution the wave field does not exactly satisfy the linear Boussinesq equations or the linear dispersion relation.  However, the two linear terms in $\disp A = \hlap A - \alpha \L A$ are the largest terms in \eqref{internalTideEqnIntro}, which means that $\disp A$ is small and the wave field almost satisfies the linear dispersion relation and that \eqref{internalTideEqnIntro} is linearly stiff.  Linear stiffness makes special time-integration schemes like the exponential time differencing used in section \ref{validation} useful for solving \eqref{internalTideEqnIntro} numerically.

An examination of terms in the hydrostatic wave equation \eqref{internalTideEqnIntro} refines notions of hydrostatic internal wave `advection' and `refraction'.  In the hydrostatic Boussinesq system, advection and refraction are each associated with three terms in the momentum and buoyancy equations with the form $\bar \bu \bcdot \bnabla \tilde u$ and $\tilde \bu \bcdot \bnabla \bar u$ for advection and refraction respectively, where $\tilde \bu$ and $\bar \bu$ are wave and mean velocity fields.  Yet only part of $\bar \bu \bcdot \bnabla \tilde u$, for example, is associated with $\J \left ( \psi, \wave A \right )$, which as \eqref{internalTideEqnIntro}'s advection term ensures Galilean invariance, has the fewest derivatives on $\psi$ and is the only surviving nonlinear term in the `WKB' limit where $\psi$ has much larger scales than $A$.  Meaningfully, the remaining parts of the Boussinesq advection terms cannot be distinguished from refraction terms, as they cancel and combine to produce the Jacobians on the second line of \eqref{internalTideEqnIntro}.  The `true' refraction terms that emerge from \eqref{internalTideEqnIntro}, with three derivatives on $\psi$ and one on $A$, are
\beq
\J \left ( A , \disp \psi \right ) + \tfrac{\ii \sigma}{f_0} \left ( \hnabla A \bcdot \hnabla \disp \psi  - \tfrac{\alpha f_0^2}{N^2} A_z \D \psi_z \right ) \per
\label{trueRefractionTerms}
\eeq
The terms in \eqref{trueRefractionTerms} are largest when $\psi$ has much smaller scales than $A$ and are some, but not all, of the terms associated with wave advection of quasi-geostrophic vorticity and buoyancy fields.  The metamorphosis of advection and refraction terms in the Boussinesq system into three types of terms in \eqref{internalTideEqnIntro} --- advection terms with one derivative on $\psi$ and three on $A$, refraction terms with three derivatives on $\psi$ and one on $A$, and intermediate terms with two derivatives on $\psi$ and $A$ each --- is due to the derivatives that operate on the nonlinear terms in the Boussinesq equations' wave operator form \eqref{waveOperatorFormTide}.

A natural question is whether the hydrostatic wave equation can be coupled to the quasi-geostrophic equation in a two-component wave-flow model similar to the models derived by \citet{XieVanneste} and \citet{wagner2016near} for near-inertial waves.  Such a coupled model may be derived by using the leading-order expressions in \eqref{leadingOrderVelocity} to evaluate the wave contribution to potential vorticity, $q^w$, defined in equation 1.3 of \cite{WagnerYoung}.  Evaluating $q^w$ and diagnosing the nonlinear mean flows associated with hydrostatic waves may reveal important analogies between nonlinear optical phenomena associated with wave dislocations and phase singularities \citep{desyatnikov2005optical} and nonlinear internal wave evolution.  And a coupled tide-flow model could elucidate the effects that strong oceanic internal tides and tide-induced mean flows have on the energetics and evolution of quasi-geostrophic fronts and eddies, the main reservoir of oceanic kinetic energy and principal agent of oceanic isopycnal stirring.

\section*{Acknowledgements}

This work was supported by the National Science Foundation under OCE-1357047.  We thank Nico Grisouard and Jennifer MacKinnon for helpful discussions and comments on a early version of this manuscript.  

\appendix

\section{Wave operator form of the hydrostatic Boussinesq equations}
\label{waveOperatorFormAppendix}

Equations \eqref{xmom} through \eqref{cont} can be formulated in terms of a wave operator.  To obtain this we first add $\p_t$\eqref{zmom} to $\p_z N^{-2}$\eqref{buoy}, multiply by $f_0^2$, and use \eqref{cont} to find
\begin{align}
f_0^2 \left ( u_x + v_y \right ) &= \L p_t + \p_z \frac{f_0^2}{N^2} \left ( \bu \bcdot \bnabla p_z \right ) \per
\label{alternateContinuity}
\end{align}
Subtracting $\p_y$\eqref{xmom} from $\p_x$\eqref{ymom}, multiplying by $f_0$, and using \eqref{alternateContinuity} yields the vertical vorticity equation, 
\beq
f_0 \omega_t +  \L p_t = - f_0 \pnabla \bcdot \left ( \bu \bcdot \bnabla \right )\bu - \p_z \frac{f_0^2}{N^2} \left ( \bu \bcdot \bnabla p_z \right ) \com
\label{hydrostaticVerticalVorticity}
\eeq
where $\pnabla = - \p_y \bxh + \p_x \byh$.  Next, adding $\p_x$\eqref{xmom} to $\p_y$\eqref{ymom}, using \eqref{alternateContinuity}, and operating on the result with $ f_0^2 \p_t$ leads to
\beq
\p_t \big ( \p_t^2 \L + f_0^2 \hlap \big ) p + \p_z \p_t^2 \frac{f_0^2}{N^2} \left ( \bu \bcdot \bnabla p_z \right ) - f_0^3 \omega_t = - f_0^2 \p_t \p_x \left ( \bu \bcdot \bnabla u \right ) - f_0^2 \p_t \p_y \left ( \bu \bcdot \bnabla v \right ) \per
\label{hydrostaticDivergence}
\eeq
Adding \eqref{hydrostaticDivergence} to $ f_0^2 \eqref{hydrostaticVerticalVorticity}$ eliminates $f_0^3 \omega_t$ and yields the wave operator form of \eqref{xmom} through \eqref{cont},
\beq
\p_t \Big [ \p_t^2 \L + f_0^2 \left ( \hlap + \L \right ) \Big ] p = - f_0^2 \left ( \p_t \hnabla + f_0 \pnabla \right ) \bcdot \left ( \bu \bcdot \bnabla \right ) \bu - \p_z \frac{f_0^2}{N^2} \left ( \p_t^2 + f_0^2 \right ) \left ( \bu \bcdot \bnabla p_z \right ) \com
\label{hydrostaticWaveOperatorFormA}
\eeq
where $\hnabla = \p_x \bxh + \p_y \byh$ is the horizontal part of the gradient operator.

\section{The part of RHS in \eqref{RHSdef} proportional to $\ee^{-\ii \sigma \tt}$}
\label{findingRHS}

In this appendix we parse the right-hand side of \eqref{o1WaveOperatorForm}, or `RHS', for its part proportional to $\ee^{-\ii \sigma \tt}$.  The RHS defined in \eqref{RHSdef} is 
\beq
\text{RHS} = - f_0^2 \left ( \p_{\tt} \hnabla + f_0 \pnabla \right ) \bcdot \left ( \bu \bcdot \bnabla \right ) \bu  - \p_z \tfrac{f_0^2}{N^2} \left ( \p_{\tt}^2 + f_0^2 \right ) \left ( \bu \bcdot \bnabla p_z \right ) \per
\label{theRHSTideAppendix}
\eeq
In \eqref{theRHSTideAppendix} and hereafter we drop the subscripts `0' denoting leading-order fields for clarity.  All fields are leading-order, so that $\left ( \bu, p \right ) = \left ( \bu_0, p_0 \right )$.  

\subsection{The leading-order solution}
\label{RHSleadingOrderSolution}

The leading-order pressure is
\beq
p = f_0 \left ( \psi + \ee^{-\ii \sigma \tt} A + \ee^{\ii \sigma \tt} A^* \right ) \com
\eeq
and the velocity $\bu$ is given in \eqref{leadingOrderVelocity}.  An expression more compact than \eqref{leadingOrderVelocity} and useful for the strenuous bookkeeping that follows is
\beq
\bu = \pnabla \psi - \frac{\ee^{-\ii \sigma \tt}}{\alpha f_0} \left ( \ii \sigma \bnablad + f_0 \pnabla \right ) A + \frac{\ee^{\ii \sigma \tt}}{\alpha f_0} \left ( \ii \sigma \bnablad - f_0 \pnabla \right ) A^* \com 
\eeq
where $\pnabla = -\p_y \bxh + \p_x \byh$ and the three-component vector operator $\bnablad$ is defined
\beq
\bnablad \defn \p_x \bxh + \p_y \byh - \frac{\alpha f_0^2}{ N^2} \p_z \bzh \per
\eeq
Notice that $\bnablad$ does not commute with $\p_z$ and that $\bnabla \bcdot \bnablad = \hlap - \alpha \L = \disp$.  The first-order advective derivative is
\beq
\bu \bcdot \bnabla = \J \left ( \psi, \bullet \right ) - \frac{\ee^{-\ii \sigma \tt}}{ \alpha f_0} \big [ f_0 \J \left ( A , \bullet \right ) + \ii \sigma \bnablad A \bcdot \bnabla \big ] + \cc \per
\label{advectiveDerivativeTide}
\eeq
The horizontal divergence and vertical vorticity $\omega \defn \pnabla \bcdot \bu$ are
\beq
\hnabla \bcdot \bu = \frac{\ii \sigma}{\alpha f_0} \hlap \Big ( \ee^{\ii \sigma \tt} A^* -  \ee^{-\ii \sigma \tt} A \Big ) \com \quad \text{and} \quad \omega = \hlap \psi - \alpha^{-1} \hlap \Big ( \ee^{-\ii \sigma \tt} A + \ee^{\ii \sigma \tt} A^* \Big ) \per
\eeq
A third useful derivative quantity is
\beq
\left ( \p_{\tt} \hnabla + f_0 \pnabla \right ) \bcdot \bu = f_0 \hlap \psi - \frac{\sigma^2 + f_0^2}{\alpha f_0} \hlap \left ( \ee^{-\ii \sigma \tt} A + \ee^{\ii \sigma \tt} A^* \right ) \per
\label{Sdotu}
\eeq
The average energy density in the hydrostatic linear solution is
\begin{align}
\ew &= \half \left ( \overline{u^2} + \overline{v^2} + N^{-2} \overline{b^2} \right ) \com \\
&= \frac{1}{2} | \hnabla \psi |^2 + \frac{f_0^2}{2 N^2} \psi_z^2  + \frac{2+\alpha}{\alpha^2}  | \hnabla A |^2 + \frac{2\ii\sqrt{1+\alpha}}{\alpha^2 }\J(A^*,A) + \frac{f_0^2}{N^2} | A_z |^2 \com
\label{linearEnergyDensity}
\end{align}
and the total, integrated `wave energy' is 
\beq
\Ew \defn \int \frac{\alpha + 2}{\alpha^2} \big | \hnabla A |^2 + \frac{f_0^2}{N^2} \big | A_z |^2 \id V \per
\label{perturbationEnergy}
\eeq
The first term in \eqref{perturbationEnergy} is total wave kinetic energy and the second term is total wave potential energy.  The Jacobian contribution to $\ew$ in \eqref{linearEnergyDensity} integrates to zero and thus does not contribute to the integral quantity $\Ew$ in \eqref{perturbationEnergy}.  $\Ew$ is conserved only over short times of $O(\sigma^{-1})$ in the hydrostatic wave equation \eqref{internalTideEqnIntro}.


\subsection{Some strenuous bookkeeping}

We tackle the momentum advection term in \eqref{theRHSTideAppendix} first, which expands into
\begin{align}
\begin{split}
f_0^2 \left ( \p_{\tt} \hnabla + f_0 \pnabla \right ) \bcdot \left ( \bu \bcdot \bnabla \right ) \bu &= f_0^2 \left ( \bu \bcdot \bnabla \right ) \left ( \p_{\tt} \hnabla + f_0 \pnabla \right ) \bcdot \bu  \\
& \quad + f_0^2 \left ( \bu_{xt} - f_0 \bu_y \right ) \bcdot \bnabla u + f_0^2 \left ( \bu_{yt} + f_0 \bu_x \right ) \bcdot \bnabla v \\
& \quad + f_0^2 \bu_x \bcdot \bnabla u_t + f_0^2 \bu_y \bcdot \bnabla v_t + f_0^2 \bu_t \bcdot \bnabla \left ( u_x + v_y \right )
\end{split}
\label{firstTermsRHS}
\end{align}
Using \eqref{advectiveDerivativeTide} and multiplying by $\ee^{\ii \alpha \tt} \alpha / f_0$ yields
\beq
\begin{split}
\ee^{\ii \sigma \tt} \alpha f_0 \left ( \bu \bcdot \bnabla \right ) \left ( \p_{\tt} \hnabla + f_0 \pnabla \right ) \bcdot \bu &= - \left ( \sigma^2 +  f_0^2 \right ) \J \left ( \psi, \hlap A \right ) - f_0^2 \J \left ( A, \hlap \psi \right ) \\
& \qquad - \ii \sigma f_0 \bnablad A  \bcdot  \bnabla \hlap \psi + \cdots \com
\end{split}
\label{T1}
\eeq
where throughout this subappendix the $\cdots$ stand for terms that do not contribute to the part of RHS proportional to $\ee^{-\ii \sigma \tt}$.  The next two terms are somewhat more involved.  We eventually obtain 
\begin{align}
\begin{split}
\ee^{\ii \sigma \tt} \alpha f_0 \left ( \bu_{xt} - f_0 \bu_y \right ) \bcdot \bnabla u  &= 2 \ii \sigma f_0 \J \left ( \psi_y , A_x \right ) \\
& \qquad + \sigma^2 \bnablad A_x \bcdot \bnabla \psi_y - \ii \sigma f_0 \bnablad A_y \bcdot \bnabla \psi_y + \cdots \com
\end{split}
\label{T2}
\end{align}
and
\begin{align}
\begin{split}
\ee^{\ii \sigma \tt} \alpha f_0 \left ( \bu_{yt} + f_0 \bu_x \right ) \bcdot \bnabla v &= -2  \ii \sigma f_0 \J \left ( \psi_x , A_y \right ) \\
& \qquad - \sigma^2 \bnablad A_y \bcdot \bnabla \psi_x - \ii \sigma f_0 \bnablad A_x \bcdot \bnabla \psi_x + \cdots \per 
\end{split}
\label{T3}
\end{align}
The fourth and fifth terms in \eqref{firstTermsRHS} are
\beq
\begin{split}
\ee^{\ii \sigma \tt} \alpha f_0 \big ( \bu_x \bcdot \bnabla u_t + \bu_y \bcdot \bnabla v_t \big ) &= - \sigma^2 \J \left ( \psi_x , A_x \right ) - \sigma^2 \J \left ( \psi_y, A_y \right ) \\
 & \qquad + \ii \sigma f_0 \J \left ( \psi_y , A_x \right ) - \ii \sigma f_0 \J \left ( \psi_x , A_y \right ) \per
\end{split}
\eeq
The sixth term in \eqref{firstTermsRHS} has no part proportional to $\ee^{-\ii \sigma t}$ because both $\bu_t$ and $u_x + v_y = - w_z$ oscillate with frequency $\sigma$.  At last, the second term in \eqref{theRHSTideAppendix} is 
\begin{align}
\begin{split}
\ee^{\ii \sigma \tt} \p_z & \tfrac{\alpha f_0}{ N^2} \left ( \p_t^2 + f_0^2 \right ) \left ( \bu \bcdot \bnabla p_z \right ) \\
&= - \p_z \tfrac{\alpha^2 f_0^2}{ N^2} \Big [ f_0^2 \J \left ( \psi, A_z \right ) -  \alpha^{-1} f_0^2 \J \left ( A, \psi_z \right ) - \ii \alpha^{-1} \sigma f_0 \bnablad A \bcdot \bnabla \psi_z \Big ] + \cdots \com
\end{split} \label{T4a} \\
\begin{split}
&= - \sigma^2 \tfrac{\alpha f_0^2}{ N^2} \J \left ( \psi_z , A_z \right ) - \alpha^2 f_0^2 \J \left ( \psi, \L A \right ) - \alpha f_0^2 \J \left ( \L \psi , A \right ) \\
& \qquad + \ii \sigma f_0 \p_z \left ( \bnablad A \bcdot \tfrac{\alpha f_0^2}{ N^2} \p_z \bnabla \psi \right )  + \cdots \per 
\end{split} 
\label{T4}
\end{align}
The extra factor of $- \alpha f_0^2 $ on the right of \eqref{T4a} comes from the relation $\sigma^2 - f_0^2 = \alpha f_0^2 $.  In passing from \eqref{T4a} to \eqref{T4} we employ the Jacobian identity $\J \left ( A , \psi_z \right ) = - \J \left ( \psi_z , A \right )$, distribute the $z$-derivative, and use $\alpha + 1 = \sigma^2 / f_0^2$.  

We next collect the contributions to $\alpha \text{RHS}/ f_0$ in $\eqref{T1} + \eqref{T2} + \eqref{T3} + \eqref{T4}$ and organize them according to whether they are multiplied by $\sigma^2$, $f_0^2$, or $\ii \sigma f_0$.  We observe a cancellation within the collection
\beq
\bnablad A_x \bcdot \bnabla \psi_y - \bnablad A _y \bcdot \bnabla \psi_x - \frac{\alpha f_0^2}{N^2} \J \left ( \psi_z , A_z \right ) = - \J \left ( \psi_x , A_x \right ) - \J \left ( \psi_y , A_y \right ) \com
\eeq
which, along with the identity
\beq
\hlap \J \left ( \psi , A \right ) = \J \left ( \hlap \psi, A \right ) + \J \left ( \psi , \hlap A \right ) + 2 \J \left ( \psi_x , A_x \right ) + 2 \J \left ( \psi_y , A_y \right ) \com
\eeq
permits the simplification of terms proportional to $\sigma^2$:
\begin{align}
\begin{split}
\frac{1}{\sigma^2} T_{\sigma^2} &= - \J \left ( \psi, \hlap A \right ) - \J \left ( \psi_x , A_x \right ) - \J \left ( \psi_y , A_y \right ) \\
& \qquad + \bnablad A_x \bcdot \bnabla \psi_y - \bnablad A _y \bcdot \bnabla \psi_x - \frac{\alpha f_0^2}{ N^2} \J \left ( \psi_z , A_z \right ) \com 
\end{split} \\ 
&= - \J \left ( \psi, \hlap A \right ) - 2 \J \left ( \psi_x , A_x \right ) - 2\J \left ( \psi_y, A_y \right ) \per
\label{Tsig}
\end{align}
Next, we employ the notation $\disp = \hlap - \alpha \L$ in writing terms proportional to $f_0^2$: 
\begin{align}
\frac{1}{f_0^2} T_{f_0^2} &= - \J \left ( \psi , \hlap A \right ) + \J \left ( \hlap \psi , A \right ) - \alpha^{2} \J \left ( \psi , \L A \right ) - \alpha \J \left ( \L \psi , A \right ) \com \\
&= - \J \left ( \psi , \hlap A \right ) - \alpha^2 \J \left ( \psi, \L A \right ) + \J \left ( \disp \psi , A \right )  \com \label{Tf2}
\end{align} 
Finally, the terms proportional to $\ii \sigma f_0$ are
\begin{align}
\begin{split}
\frac{1}{\ii \sigma f_0} T_{\sigma f_0} &= 3 \J \left ( \psi_y , A_x \right ) - 3 \J \left ( \psi_x , A_y \right ) \\
& \qquad - \bnablad A \bcdot \bnabla \hlap \psi - \bnablad A_x \bcdot \bnabla \psi_x - \bnablad A_y \bcdot \bnabla \psi_y \\
& \qquad + \p_z \left ( \bnablad A \bcdot \tfrac{\alpha f_0^2}{ N^2} \p_z \bnabla \psi \right ) \com 
\end{split}
\label{sigeff1}
\end{align}
Some rearrangement and combination of terms leads eventually to the identity
\beq
\begin{split}
\bnablad A \bcdot \bnabla \hlap \psi &+ \bnablad A_x \bcdot \bnabla \psi_x + \bnablad A_y \bcdot \bnabla \psi_y - \p_z \left ( \bnablad A \bcdot \tfrac{\alpha f_0^2}{ N^2} \p_z \bnabla \psi \right ) \\
&= \J \left ( \psi_y, A_x \right ) - \J \left ( \psi_x, A_y \right ) + \p_x \left ( A_x \disp \psi \right ) + \p_y \left ( A_y \disp \psi \right ) \\
& \qquad - \disp \left ( \tfrac{\alpha f_0^2}{N^2} \psi_z A_z \right ) + \p_z \left ( \tfrac{\alpha f_0^2}{N^2} \psi_z \disp A \right ) \per
\end{split}
\label{strenuousIdentity}
\eeq
Using \eqref{strenuousIdentity} to simplify \eqref{sigeff1} yields
\beq
\begin{split}
\frac{1}{\ii \sigma f_0} T_{\sigma f_0} &= 2 \J \left ( \psi_y , A_x \right ) - 2 \J \left ( \psi_x , A_y \right ) - \hnabla \bcdot \left ( \disp \psi \hnabla A \right )  \\
& \qquad  + \disp \left ( \tfrac{\alpha f_0^2}{N^2} \psi_z A_z \right ) - \p_z \left ( \tfrac{\alpha f_0^2}{N^2} \psi_z \disp A \right ) \per
\end{split}
\label{sigeff2}
\eeq

\subsection{The final tally}

With \eqref{Tsig}, \eqref{Tf2}, and \eqref{sigeff2}, we have all the pieces needed to construct  RHS.  We find that 
\begin{align}
\tfrac{\alpha}{f_0} \overline{ \ee^{\ii \sigma \tt} \text{RHS} } &= - \left ( T_{\sigma^2} + T_{f_0^2} + T_{\sigma f_0} \right ) \com \\
\begin{split}
&= \left ( \sigma^2 + f_0^2 \right ) \J \left ( \psi, \hlap A \right ) + \left ( \alpha f_0 \right )^2 \J \left ( \psi, \L A \right ) - f_0^2 \J \left ( \disp \psi, A \right ) \\
& \quad - 2 \ii \sigma \Big [  \J \left ( \psi_x , \ii \sigma A_x - f_0 A_y \right ) +  \J \left ( \psi_y , \ii \sigma A_y + f_0 A_x \right ) \Big ]  \\
& \quad  + \ii \sigma f_0 \left [ \hnabla \bcdot \left ( \disp \psi \hnabla A \right ) - \disp \left ( \tfrac{\alpha f_0^2}{N^2} \psi_z A_z \right ) +  \p_z \left ( \tfrac{\alpha f_0^2}{N^2} \psi_z \disp A \right )   \right ] \per
\end{split}
\end{align}


\bibliographystyle{jfm}
\bibliography{refs}

\end{document}